\documentclass[structabstract]{aa}  
\usepackage{graphicx}
\usepackage{subfigure}
\usepackage{longtable}
\usepackage{footmisc}
\usepackage{amsmath}
\usepackage{txfonts}
\usepackage{natbib}
\bibpunct{(}{)}{;}{a}{}{,}
\begin{document}
   \title{Kuiper Belts Around Nearby Stars\thanks{Based on observations with APEX, Llano Chajnantor, Chile (OSO programme 081.F-9330(A)).}}


   \author{R. Nilsson
          \inst{1}
          \and
           R. Liseau
          \inst{2}
          \and
           A. Brandeker
          \inst{1}
          \and
           G. Olofsson
          \inst{1}
	 \and
	  G. L. Pilbratt
	 \inst{3}
	 \and
	 C. Risacher
	 \inst{4}
          \and
           J. Rodmann
          \inst{5}  
          \and
	 J.-C. Augereau
	\inst{6}
	\and
          P. Bergman
          \inst{2}
          \and
          C. Eiroa
          \inst{7}
	 \and
          M. Fridlund
          \inst{3}
	 \and
          P. Th\'ebault
          \inst{8,1}
	 \and
	 G. J. White
	\inst{9,10}
          }
          
   \offprints{R. Nilsson}

   \institute{Department of Astronomy, Stockholm University, AlbaNova University Center, Roslagstullsbacken 21, SE-106 91 Stockholm, Sweden\\
              \email{ricky@astro.su.se, alexis@astro.su.se, olofsson@astro.su.se}
         \and
             Onsala Space Observatory, Chalmers University of Technology, SE-439 92 Onsala, Sweden\\
             \email{rene.liseau@chalmers.se, per.bergman@chalmers.se}
	\and
             ESA Astrophysics Missions Division, ESTEC, PO Box 299, 2200 AG Noordwijk, The Netherlands\\
             \email{malcolm.fridlund@esa.int, gpilbratt@rssd.esa.int}
         \and
            SRON, Postbus 800, 9700 AV Groningen, The Netherlands\\
             \email{crisache@sron.nl}
         \and
	   ESA/ESTEC Space Environment and Effects Section, PO Box 299, 2200 AG Noordwijk, The Netherlands\\
	\email{jens.rodmann@esa.int}
	\and
             Universit\'e Joseph Fourier/CNRS, Laboratoire d'Astrophysique de Grenoble, UMR 5571, Grenoble, France\\
             \email{augereau@obs.ujf-grenoble.fr}
	\and
             Dpto. F\' isica Te\'orica, Facultad de Ciencias, Universidad Aut\'onoma de Madrid, 28049 Madrid, Spain\\
             \email{carlos.eiroa@uam.es}
	\and
	   Observatoire de Paris, Section de Meudon, F-92195 Meudon Principal Cedex, France\\
	   \email{philippe.thebault@obspm.fr}
	\and
             Department of Physics and Astronomy, Open University, Walton Hall, Milton Keynes MK7 6AA, UK\\
             \email{g.j.white@open.ac.uk}
	\and
             Science and Technology Facilities Council, Rutherford Appleton Laboratory, Chilton, Didcot OX11 0QX, UK\\
         		 }

   \date{Received 17 March 2010 / 
   				Accepted May 2010}

 
  \abstract
   {The existence of dusty debris disks around a large fraction of solar type main-sequence stars, inferred from excess far-IR and submillimetre emission compared to that expected from stellar photospheres, suggests that leftover planetesimal belts analogous to the asteroid- and comet reservoirs of the Solar System are common.}
   {In order to detect and characterise cold extended dust originating from collisions of small bodies in disks, belts, or rings at Kuiper-Belt distances (30--50\,AU or beyond) sensitive submillimetre observations are essential. Measurements of the flux densities at these wavelengths will extend existing IR photometry and permit more detailed modelling of the Rayleigh-Jeans tail of the disks spectral energy distribution (SED), effectively constraining dust properties and disk extensions. By observing stars spanning from a few up to several hundred Myr, the evolution of debris disks during crucial phases of planet formation can be studied.}
   {We have performed 870\,$\mu$m observations of 22 exo-Kuiper-Belt candidates, as part of a Large Programme with the LABOCA bolometer at the APEX telescope. Dust masses (or upper limits) were calculated from integrated 870\,$\mu$m fluxes, and fits to the SED of detected sources revealed the fractional dust luminosities $f_{\mathrm{dust}}$, dust temperatures $T_{\mathrm{dust}}$, and power-law exponents $\beta$ of the opacity law.}
   {A total of 10 detections with at least 3$\sigma$ significance were made, out of which five (HD\,95086, HD\,131835, HD\,161868, HD\,170773, and HD\,207129) have previously never been detected at submillimetre wavelengths. Three additional sources are marginally detected with $>2.5\sigma$ significance. The best-fit $\beta$ parameters all lie between 0.1 and 0.8, in agreement with previous results indicating the presence of grains that are significantly larger than those in the ISM. From our relatively small sample we estimate $f_{\mathrm{dust}}$\,$\propto$\,$t^{-\alpha}$, with $\alpha$\,$\sim$\,0.8--2.0, and identify an evolution of the characteristic radial dust distance $R_{\mathrm{dust}}$ that is consistent with the $t^{1/3}$ increase predicted from models of self-stirred collisions in debris disks.}
   {}

	 \titlerunning{Kuiper Belts Around Nearby Stars}
   \maketitle
%

\section{Introduction}
The existence of warm dust around solar type main-sequence stars is implied from infrared (IR) excess emission (above the stellar photospheric contribution) discovered by the \emph{Infrared Astronomical Satellite} \citep[IRAS, e.g.][]{Aumann1984,Neugebauer1984}, and studied by the \emph{Infrared Space Observatory} \citep[ISO, e.g.][]{Kessler1996}, and \emph{Spitzer} \citep[e.g.][]{Werner2004}. Since such dust has a limited lifetime its existence suggests the presence of larger asteroidal and/or cometary bodies which through collisions continuously replenish the dusty debris disk. Excess emission at wavelengths of 25\,{$\mu$}m and shorter comes predominantly from hot dust ($>$300\,K), located within 1--5\,AU of the star (similar to the asteroidal belt in our Solar system), while an excess at 25--100\,{$\mu$}m probes cooler dust ($\sim$100--300\,K) located at 5--10\,AU. However, in order to detect cold and very extended dust, originating in Kuiper-Belt analogs (on the order of 100\,AU from the star), observations at submillimetre (submm) wavelengths are needed. Although very few older stars show evidence of hot dust, likely due to clearing by planets \citep[e.g.][]{Beichman2006a,Meyer2008}, cool debris disks appear more common and many should be detectable at submm wavelengths based on extrapolation from observed far-IR fluxes.

The potential correlation of cold dust and planetesimal belts with the formation of exo-planets in these systems is another incentive for larger submm surveys of nearby solar-type main-sequence stars. \citet{Beichman2006b} and \citet{Bryden2006} reported \emph{Spitzer} MIPS observations of a large sample of nearby F, G, and K main-sequence stars, and detected an overall excess of 70\,{$\mu$}m emission towards 13\% of the stars. From a larger sample of solar-type stars, spanning ages between 3\,Myr and 3\,Gyr, studied in the \emph{Spitzer} Legacy programme Formation and Evolution of Planetary Systems (FEPS), showed a total 70\,{$\mu$}m excess rate of 7\% with high (50 times the photosphere) fractional excesses at ages between 30 and 200\,Myr, and significantly lower fractional excess for older stars \citep{Carpenter2009,Hillenbrand2008a}. There was no clear trend for the temporal evolution of 70\,{$\mu$}m excess in FEPS, however \citet{Su2006} estimate a decay time of $\sim$400\,Myr for A type stars. At submm and mm wavelengths 3\% of the FEPS stars show evidence of continuum emission, with a suggested decrease in dust masses and/or changes in the grain properties for stellar ages of 10--30\,Myr \citep{Carpenter2005}.

An additional advantage in making observations in the submm region is that measured integrated fluxes are directly proportional to the temperature and mass of the disk, due to the fact that they (in most cases) sample the Rayleigh-Jeans tail of the spectral energy distribution (SED), and that the disk can be assumed to be optically thin at these wavelengths. At the same time, the thermal radiation in the submm is dominated by large and cool dust grains, thus providing a better estimate of the total dust mass (although still giving just a lower limit). By combining submm observations with IR-photometry a fit to the SED can be made, yielding constraints on the temperature and radial extent of spatially unresolved disks.

In this paper we present the first results of a Large Programme submm-survey with the \emph{Large APEX Bolometer CAmera} (LABOCA) at the \emph{Atacama Pathfinder EXperiment} (APEX) telescope, targeting stars of spectral type B to M, belonging to nearby stellar associations with ages ranging between 10 and 100\,Myr. While the main aim of the study is to investigate disk evolution by relating disk parameters and grain properties to fundamental stellar parameters (metallicity, mass, and age), we also hope to spatially resolve extended dust disks around some of the most nearby stars. The observational scheme is designed to reach $\sim$2\,mJy/beam root-mean-square (RMS) sensitivity, enabling the detection of debris disk masses as low as the mass of the Moon \citep[comparable to the Solar System's Kuiper-Belt mass of 3\,$M_{\mathrm{Moon}}$;][and references therein]{Iorio2007}, and to make observations of exo-Kuiper-Belts out to distances of about 50--100\,pc. A successful precursor study of the $\sim$12\,Myr-old $\beta$ Pictoris Moving Group clearly detected 2 out of 7 stars \citep{Nilsson2009}.

\section{Stellar Sample}\label{sec:sample}
The sample of 22 F-, G-, and K-type stars selected for this first round of observations were based on levels of far-IR excess (measured by \emph{IRAS}, \emph{ISO} and \emph{Spitzer}) and proximity (also making them targets for \emph{Herschel Space Observatory} Key Programmes on debris disks), in order to optimise initial detection rates and get meaningful upper limits. Positions, spectral types, ages, and distances of the objects are listed in Table\,\ref{table:1}, together with integration times of the observations. The age estimates of most stars, which are based on lithium abundances, H-R diagram location, moving group (MG) association, chromospheric activity, etc., all carry large uncertainties. E.g., HD\,105 is a member of the Tuc-Hor MG \citep{Mamajek2004}, with an estimated age of 27$\pm$11\,Myr \citep[][]{Mentuch2008}, but could be as old as 225\,Myr based on the level of chromospheric activity \citep{Apai2008}; and HD\,98800, HD\,25457,  HD\,141569, HD\,152404, and HD\,207129 might still be in their pre-main-sequence phase. Due to these inconsistencies we have adopted probable approximate ages with upper and lower limits that encompass the range of published values from different methods. Consequently, our ability to resolve details about the temporal evolution of disk properties in the young 10--100\,Myr systems, where terrestrial planet formation could still be in progress, according to, e.g., isotopic dating of meteorites and the Earth-Moon system \citep{Touboul2007} and dynamical modelling \citep{Mandell2007}, is somewhat restricted. 



\section{Observations and Data Reduction}
The 12-m diameter submm telescope APEX \citep{Gusten2006} is located at an altitude of 5100\,m in the Chilean Andes, offering excellent atmospheric transparency. It is operated jointly by Onsala Space Observatory, the Max-Planck-Institut f{\"u}r Radioastronomie, and the European Southern Observatory. Since 2007, the LABOCA bolometer array has been available for observations at a central wavelength of 870\,{$\mu$}m (345\,GHz) and bandwidth of 150\,{$\mu$}m (60\,GHz), covering a 11$\farcm$4 field-of-view with its array of 295 bolometers; each with a nearly circular 19$\farcs$2 full-width at half-maximum (FWHM) Gaussian beam \citep{Siringo2009}. A spiral pattern mapping mode was employed in order to recover fully sampled maps from the under-sampled bolometer array. \citet{Siringo2009} provide a detailed description of the instrument, observing modes, performance and sensitivity.

Observations of the targets presented in Table \ref{table:1} were obtained between June 4 and October 22, 2008, amounting to roughly 80 out of 200 allocated hours of this Large Programme. Several individual 7.5\,min long spiral scans, producing raw maps with an approximately uniform noise distribution within a radius of about 4$\arcmin$ around the central position of the source, were obtained for each star. In-between these scans, skydips (to determine the correction for the atmospheric opacity), flux calibration, and measurements of pointing accuracy and focussing were performed on selected calibration objects.

\begin{table*}
\caption{\label{table:1} Observing log with stellar properties and integration times.}
\centering
\begin{tabular}{l l c c c c c c c}
\hline\hline
Object ID & Name & R.A.\tablefootmark{a} & Dec. & Spectral Type\tablefootmark{b} & Age$^{\mathrm{upper}}_{\mathrm{lower}}$ & Reference & Distance\tablefootmark{c} & Integration time\\
 &  & (h m s) & ($\degr$ $\arcmin$ $\arcsec$) &  & (Myr) &  & (pc) & (min) \\
\hline
HD\,105 &  & 00 05 52.54 & -51 45 11.0 & G0V & $30^{3500}_{27}$ & 1, 2, 27 & 40 & 104 \\

HD\,17390 &  & 02 46 45.10 & -21 38 22.3 & F3IV/V & $300^{800}_{?}$ & 3, 4 & 45 & 24 \\

HD\,21997 &  & 03 31 53.65 & -25 36 50.9 & A3IV/V & $50^{150}_{200}$ & 5, 3, 6 & 74 & 52 \\

HD\,25457 &  & 04 02 36.74 & -00 16 08.1 & F5V\tablefootmark{d} & $50^{150}_{30}$ & 7 & 19 & 97 \\

HD\,30447 &  & 04 46 49.53 & -26 18 08.8 & F3V & $20^{100}_{10}$ & 6, 8 & 78 & 65 \\

HD\,31392 &  & 04 54 04.21 & -35 24 16.3 & K0V & $1300^{4000}_{400}$ & 9 & 26 & 53 \\

HD\,32297 &  & 05 02 27.44 & +07 27 39.7 & A0V & $30^{50}_{20}$ & 10 & 112 & 39 \\

HD\,61005 &  & 07 35 47.46 & -32 12 14.0 & G8V & $180^{320}_{100}$ & 11 & 35 & 53 \\

HD\,78702 &  & 09 09 04.21 & -18 19 42.8 & A0/A1V & $220^{320}_{80}$ & 6 & 80 & 91 \\

HD\,95086 &  & 10 57 03.02 & -68 40 02.4 & A8III & $20^{50}_{15}$ & 12, 13, 14 & 92 & 32 \\

HD\,98800 & TV\,Crt & 11 22 05.29 & -24 46 39.8 & K5V+K7V+M1V+?\tablefootmark{e} & $10^{20}_{3}$ & 15 & 47 & 62 \\

HD\,109573 & HR\,4796 & 12 36 01.03 & -39 52 10.22 & A0V+M2.5\tablefootmark{f} & $10^{20}_{4}$ & 16, 17, 18 & 67 & 185 \\

HD\,131835 &  & 14 56 54.47 & -35 41 43.6 & A2IV & $14^{18}_{10}$ & 8, 6 & 111 & 145 \\

HD\,139664 &  & 15 41 11.38 & -44 39 40.3 & F4V & $200^{300}_{150}$ & 7 & 18 & 170 \\

HD\,141569 &  & 15 49 57.75 & -03 55 16.4 & A0Ve+M2+M4\tablefootmark{g} & $5^{8}_{2}$ & 19, 26 & 99 & 206 \\

HD\,152404 & AK\,Sco & 16 54 44.85 & -36 53 18.6 & F5V+F5V\tablefootmark{h} & $17^{18}_{16}$ & 20 & 145 & 127 \\

HD\,161868 & $\gamma$\,Oph & 17 47 53.56 & +02 42 26.2 & A0V & $180^{190}_{170}$ & 16, 17 & 29 & 218 \\

HD\,164249 &  & 18 03 03.41 & -51 38 56.4 & F5V & $12^{20}_{8}$ & 21 & 47 & 166 \\

HD\,170773 &  & 18 33 00.92 & -39 53 31.3 & F5V & $200^{300}_{150}$ & 3, 5 & 36 & 311 \\

HD\,193307 &  & 20 21 41.03 & -49 59 57.9 & G0V & $5700^{6300}_{5100}$ & 22, 27 & 32 & 90 \\

HD\,195627 &  & 20 35 34.85 & -60 34 54.3 & F0V & $30^{230}_{10}$ & 3, 12 & 28 & 98 \\

HD\,207129 &  & 21 48 15.75 & -47 18 13.0 & G2V & $3000^{8000}_{10}$ & 5, 22, 24, 25, 27 & 16 & 354 \\
\hline
\end{tabular}
\tablebib{
(1)~\citet[][]{Hollenbach2005} and references therein; (2) \citet{Apai2008}; (3) \citet{Zuckerman2004b}; (4) \citet{Zuckerman2004a};
(5) \citet{Song2003}; (6) \citet{Moor2006}; (7) \citet[][]{Lopez2006} and references therein; (8) \citet[][]{Rhee2007} and references therein;
(9) \citet[][]{Carpenter2009} and references therein; (10) \citet{Perryman1997}; (11) \citet{Roccatagliata2009}; (12) \citet{Lowrance2000};
(13) \citet{deZeeuw1999}; (14) \citet{Mamajek2002}; (15) \citet{Barrado2006}; (16) \citet{Song2001}; (17) \citet{Rieke2005};
(18) \citet{Stauffer1995}; (19) \citet{Weinberger2000}; (20) \citet[][]{Chen2005b} and references therein; (21) \citet{Zuckerman2001a};
(22) \citet{Mamajek2008}; (23) \citet{Henry1996}; (24) \citet[][]{Bryden2006} and references therein; (25) \citet{Zuckerman2000};
(26) \citet{Merin2004}; (27) \citet{Valenti2005}.
}
\tablefoot{
\tablefoottext{a}{Positions are from the Hipparcos Catalogue \citep{Perryman1997}.}
\tablefoottext{b}{Spectral types are from \citet{Torres2006,Thoren2000,Manoj2006,Thevenin1999}.}
\tablefoottext{c}{Distances are from the Hipparcos Catalogue \citep{Perryman1997}}
\tablefoottext{d}{From \citet{Lopez2006}. Marked as T Tau-type in SIMBAD astronomical database.}
\tablefoottext{e}{Quadruple system composed of two spectroscopic binaries, Aa+b and Ba+b, separated by 0$\farcs$848 \citep{Prato2001}. IR excess originates from the Ba+b (K7+M1V binary) component (for which the coordinates are given) \citep[see][and references therein]{Laskar2009}.}
\tablefoottext{f}{Binary system with the secondary (M2.5 pre-main-sequence) star located 7$\farcs$7 to the southwest of the primary (A0V). IR excess comes from HD\,109573A (also designated HR\,4796A or GQ\,Lup) with an imaged circumstellar ring \citep{Schneider2009}.}
\tablefoottext{g}{Primary component HD\,141569A, which is possibly still in its pre-main-sequence stage, has a binary companion consisting of two M stars at $\sim8\arcsec$ separation. The double pair could be unbound and merely flying by \citep{Reche2009}.}
\tablefoottext{h}{Spectroscopic binary with high IR excess attributed to long-lived primordial, circumbinary disk \citep{Uzpen2007,Gomez2009}.}
}
\end{table*}

The scans were reduced and combined using MiniCRUSH (v1.05), an adapted version of CRUSH \citep[Comprehensive Reduction Utility for SHARC-2,][]{Kovacs2008}, with further processing for analysis and plotting of final maps in Matlab (v7.9.0). As all of the sources were relatively faint we used filtering and settings optimised for point source extraction, with full beam smoothing, resulting in clean maps with $27\arcsec$ effective resolution. Details of the data reduction procedure can be found in \citet{Kovacs2008} and are summarised in \citet{Nilsson2009}. The general reduction steps involve flux calibration (opacity correction and counts-to-Jy conversion), flagging of bad (unresponsive, very noisy, or those that were too fast or too slow) channels, correlated noise removal, despiking, data weighting, and map making. \citet{Siringo2009} can be consulted for a description of these steps and their relation to the observing mode of the instrument.

The final maps had a sensitivity between 2 and 10\,mJy/beam. The integrated flux density of detected sources was found by fitting a 2-D Gaussian to the source region after baseline subtraction and gradient correction, with errors estimated from root-mean-square (RMS) noise calculation in the specific integration region, together with an absolute calibration error estimated to be 10\% \citep{Siringo2009}.

\section{Results}
Of the 22 far-IR excess stars observed, 10 (HD\,21997, HD\,95086, HD\,98800, HD\,109573, HD\,131835, HD\,141569, HD\,152404, HD\,161868, HD\,170773, and HD\,207129) were detected with at least a 3$\sigma$ significance, giving a detection rate of $\sim$45\%. Five of the debris disks (HD\,95086, HD\,131835, HD\,161868, HD\,170773, and HD\,207129) have previously never been seen at submm wavelengths. HD\,105, HD\,30447, and HD\,195627 are listed as marginally detected in Table\,\ref{table:2}, showing flux peaks of $>2.5\sigma$ significance at the position of the star.

\begin{table*}
\caption{\label{table:2} Integrated flux density, root-mean-square noise levels, and derived dust temperature, power-law exponent of the opacity law, mass, fractional dust luminosity, and characteristic radial dust distances for the 22 stars observed at 870-$\mu$m.}
\centering
\begin{tabular}{l c c c c c c c}
\hline\hline
Object ID & Integrated Flux Density, $F$ & RMS Noise, $\sigma$ & $T_{\mathrm{dust}}$\tablefootmark{a} & $\beta$ & Dust Mass\tablefootmark{b}, $M_\mathrm{dust}$ & $f_\mathrm{dust}$ & $R_\mathrm{dust}$\\
 & (mJy) & (mJy/beam) & (K) &  & ($M_{\mathrm{Moon}}$) & ($10^{-4}$) & (AU)\\
\hline
\multicolumn{8}{c}{DETECTIONS ($>3\sigma$)} \\
\hline
HD\,21997 & 17.6$\pm$8.0 & 4.6  & 52 & 0.7 & 33$\pm$15 & 4.9 & 240 \\ 

HD\,95086 & 41.3$\pm$18.4 & 10.4 & 150+(15--45)\tablefootmark{c} & ... & $>$76\tablefootmark{d} & 6.4--14 & 16+(250--1500) \\ 

HD\,98800 & 33.6$\pm$8.4 & 3.9 & 155\tablefootmark{e} & 0.1 & 8.5 & 470 & 2.4 \\ 

HD\,109573 & 21.5$\pm$6.6 & 3.1 & 88\tablefootmark{f} & 0.7 & 19.5$\pm$6.0 & 38 & 77 \\ 

HD\,131835 & 8.5$\pm$4.4 & 3.0 & 89 & 0.1 & 21$\pm$11 & 25 & 47 \\ 

HD\,141569 & 12.6$\pm$4.6 & 2.3 & 80\tablefootmark{g} & 0.8 & 27$\pm$10 & 73 & 110 \\

HD\,152404 & 42.9$\pm$9.8 & 3.7 & ...\tablefootmark{h} & ... & ... & ... & ... \\

HD\,161868 & 12.8$\pm$5.2 & 2.5 & 73 & 0.5 & 2.6$\pm$1.1 & 0.90 & 120 \\ 

HD\,170773 & 18.0$\pm$5.4 & 2.6 & 43 & 0.7 & 9.7$\pm$2.9 & 4.5 & 170 \\ 

HD\,207129 & 5.1$\pm$2.7 & 2.1 & 38 & 0.7 & 0.6$\pm$0.3 & 0.94 & 130 \\ 
\hline
\multicolumn{8}{c}{MARGINAL DETECTIONS ($>2.5\sigma$)} \\
\hline
HD\,105 & 10.7$\pm$5.9 & 3.7 & 40 & 0.6 & 7.7$\pm$4.2 & 2.6 & 110 \\ 

HD\,30447 & 6.9$\pm$5.0 & 4.5 & 61 & 0.3 & 12.3$\pm$8.9 & 9.3 & 56 \\ 

HD\,195627 & 13.0$\pm$7.1 & 4.1 & 57 & 0.3 & 3.1$\pm$1.7 & 1.0 & 74 \\ 
\hline
\multicolumn{8}{c}{NON-DETECTIONS} \\
\hline
HD\,17390 & ... & 6.8 & 55\tablefootmark{i} & ... & $<$13.4 & 1.9\tablefootmark{j} & ... \\

HD\,25457 & ... & 3.3 & 70\tablefootmark{k} & ... & $<$0.9 & 1.0\tablefootmark{k} & ... \\ 

HD\,31392 & ... & 4.7 & 49\tablefootmark{k} & ... & $<$3.4 & 1.6\tablefootmark{k} & ... \\

HD\,32297 & ... & 6.5 & 30\tablefootmark{l} & ... & $<$150 & 33\tablefootmark{j} & ... \\ 

HD\,61005 & ... & 6.0 & 58\tablefootmark{k} & ... & $<$6.6 & 25\tablefootmark{k} & ... \\

HD\,78702 & ... & 4.2 & 36\tablefootmark{m} & 2.0\tablefootmark{m} & $<$39 & 2.6\tablefootmark{m} & ... \\ 

HD\,139664 & ... & 3.5 & 78\tablefootmark{n} & ... & $<$0.7 & 1.2\tablefootmark{n} & ... \\ 

HD\,164249 & ... & 3.5 & 78 & ... & $<$5.3 & 5.9\tablefootmark{o} & ... \\

HD\,193307\tablefootmark{p} & ... & 3.8 & ... & ... & ... & ... & ...\\
\hline
\end{tabular}
\tablefoot{
\tablefoottext{a}{For detected and marginally detected sources we use the dust temperature derived from the best fit to the spectral energy distribution.}
\tablefoottext{b}{Upper $3\sigma$ limit on the dust mass for undetected sources.}
\tablefoottext{c}{In order not to be resolved the second dust belt should be at $<$1500\,AU, which would mean $>$15\,K, while a temperature $>$45\,K would not be compatible with far-IR data.}
\tablefoottext{d}{Calculated for upper limit on dust temperature.}
\tablefoottext{e}{Ambiguous fit (see discussion in Section\,\ref{sec:sed}).}
\tablefoottext{f}{Source appears spatially resolved (see discussion in Appendix~\ref{sec:appena}). See also more detailed modelling by \citet{Augereau1999} and SCUBA measurement by \citet{Sheret2004}.}
\tablefoottext{g}{Source appears spatially resolved (see discussion in Appendix~\ref{sec:appena}). See also modelling by \citet{Sheret2004}.}
\tablefoottext{h}{No simple debris disk model can be fitted (see Section\,\ref{sec:sed}).}
\tablefoottext{i}{From \citet{Rhee2007}.}
\tablefoottext{j}{From \citet{Moor2006}.}
\tablefoottext{k}{From \citet{Hillenbrand2008a}.}
\tablefoottext{l}{From \citet{Maness2008}.}
\tablefoottext{m}{From \citet{Williams2006}.}
\tablefoottext{n}{From \citet{Beichman2006b}.}
\tablefoottext{o}{From \citet{Rebull2008}.}
\tablefoottext{p}{Not enough photometry data for reliable SED fit.}
}
\end{table*}

\begin{figure*}
\centering
     \subfigure[]{
          \label{fig:A1m}
          \includegraphics[width=.45\textwidth]{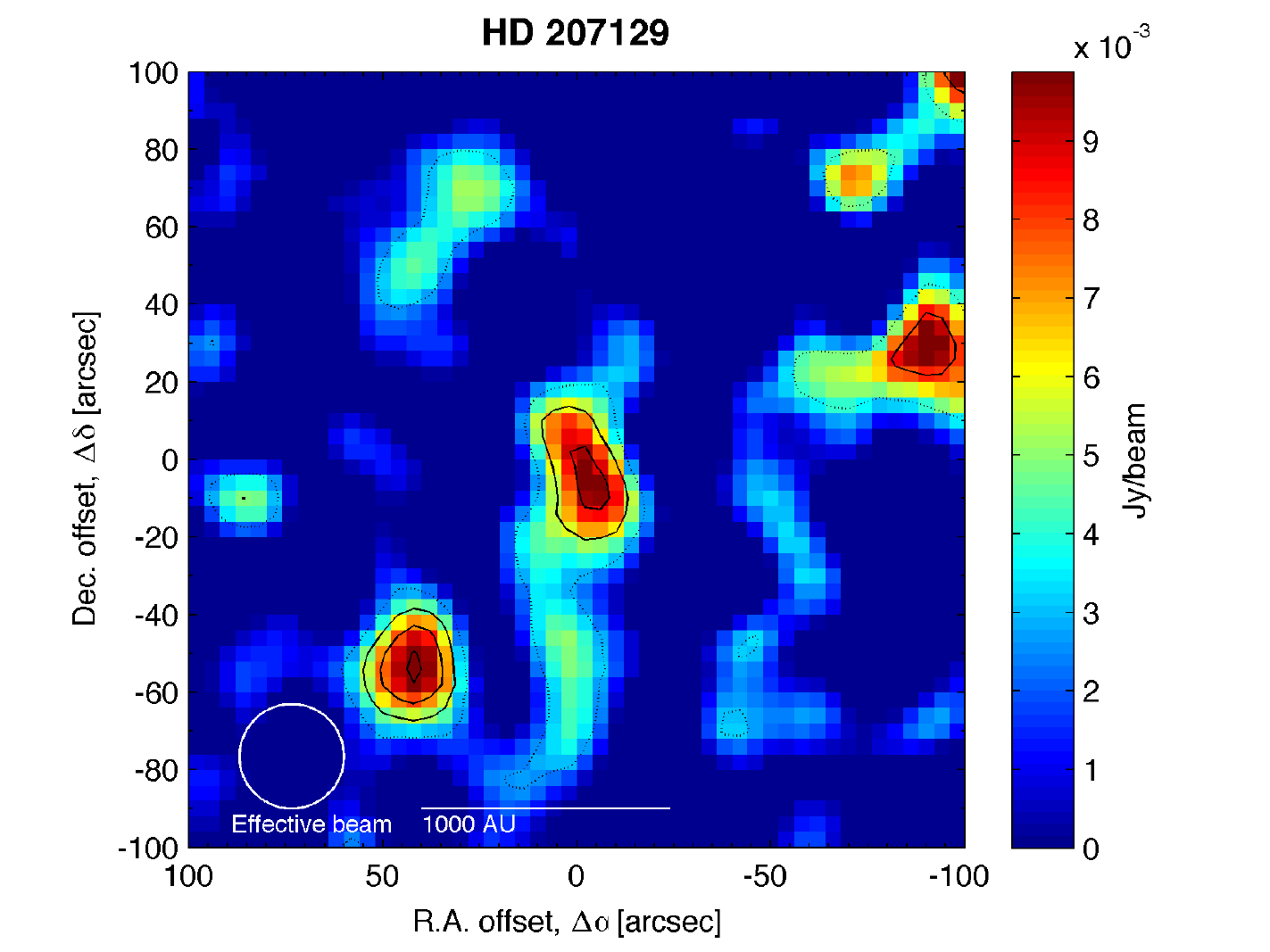}} \\
     \caption{Maps of objects detected at 870\,$\mu$m with LABOCA at the APEX telescope. Each map has been smoothed with a circular Gaussian corresponding to the FWHM of the beam, and the resulting effective FWHM image resolution of 27{$\arcsec$} is shown as a circle in the maps. In addition, a scale bar corresponding to 1000\,AU at the distance of the star is inserted. The first solid contour represents 2$\sigma$ flux levels, with the following contours at increments of 1$\sigma$ (or 2$\sigma$ for HD\,98800 and HD\,152404), while the dotted contour outlines the 1$\sigma$ level. The position of the star is at $(\Delta\alpha,\Delta\delta)=(0\arcsec,0\arcsec)$.}
     \label{fig:A1all}
\end{figure*}

The final 870\,$\mu$m flux-density maps of the detected and marginally detected objects are found in Fig.\,\ref{fig:A1all}, which show the flux level color-coded in a scale ranging from zero to the peak flux density (in Jy/beam) of each individual map. The dotted contour represents the 1$\sigma$-level and the first solid contour outlines the 2$\sigma$-level, with subsequent contour lines spaced by 1$\sigma$. A circle representing the effective $27\arcsec$ beam size after smoothing, and a line showing the angular size subtended by 1000\,AU at the distance of the object, has been inserted in the lower left corner.

Several significant flux density peaks can be found surrounding the position of the observed star, also in parts of the LABOCA field-of-view outside the plotted $200\times200\arcsec$ maps. Considering the (down to) 2\,mJy/beam sensitivity we are expected to hit the extragalactic confusion limit, permitting detection of many background submm galaxies \citep[see e.g.,][]{Wang2004}. This also effects the interpretation of our results for HD\,109573 and HD\,141569, which appear spatially resolved even with our rather large effective beam which had been optimised for point source extraction in the reductions. These issues are discussed further in Appendix~\ref{sec:appena}.

The measured integrated fluxes and RMS noise levels are listed in Table\,\ref{table:2}, together with the disk parameters (temperature, power-law exponent of the opacity law, dust mass, fractional dust luminosity, and characteristic radial distance of the dust) derived from the analysis outlined in Section\,\ref{sec:dis}. For undetected sources, an upper 3$\sigma$ limit on the dust mass is given.

\begin{figure*}
     \centering
     \subfigure[]{
          \label{fig:A1a}
          \includegraphics[width=.45\textwidth]{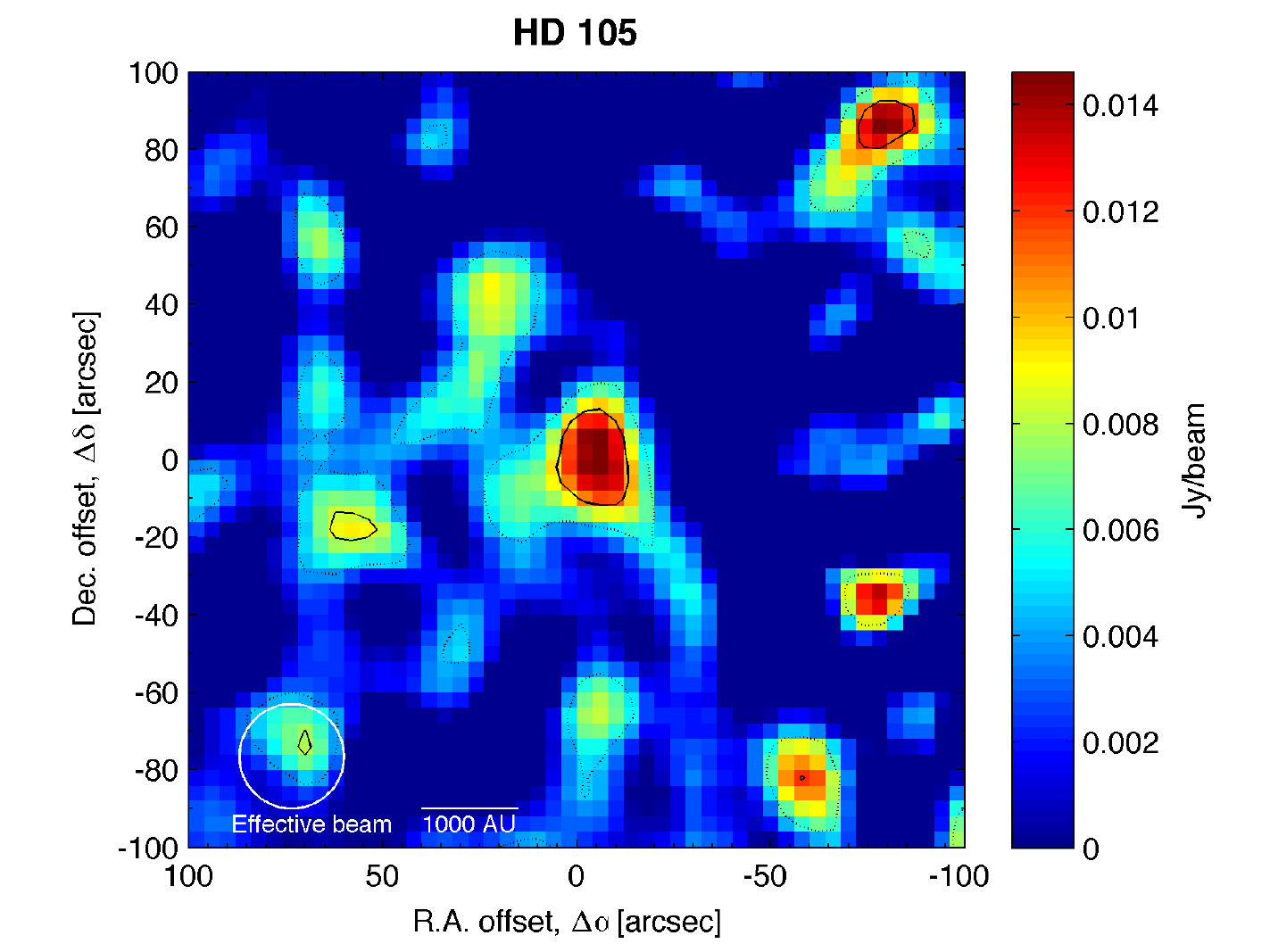}}
     \hspace{1mm}
     \subfigure[]{
          \label{fig:A1b}
          \includegraphics[width=.45\textwidth]{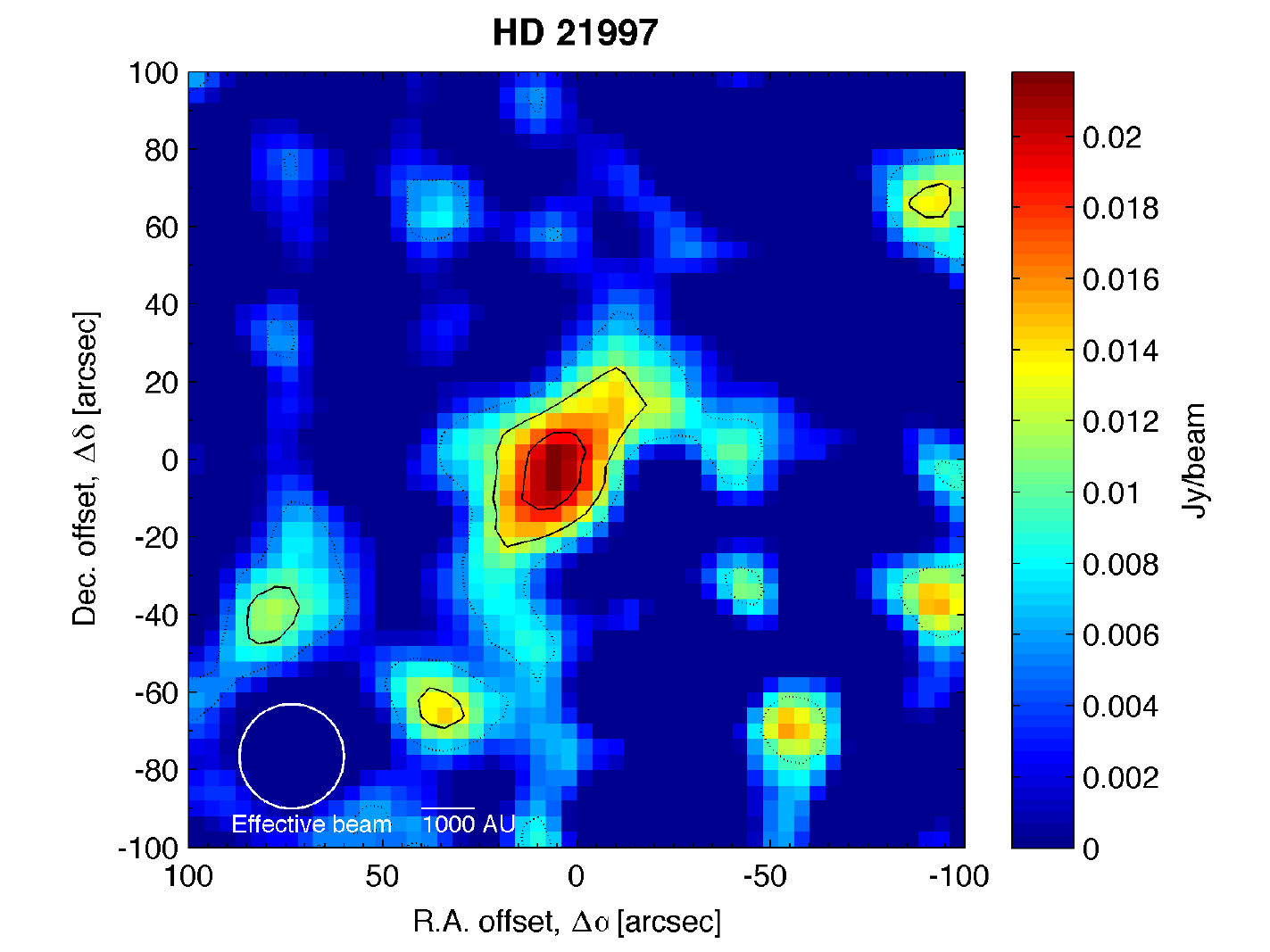}} \\
     \vspace{1mm}
     \subfigure[]{
           \label{fig:A1c}
          \includegraphics[width=.45\textwidth]{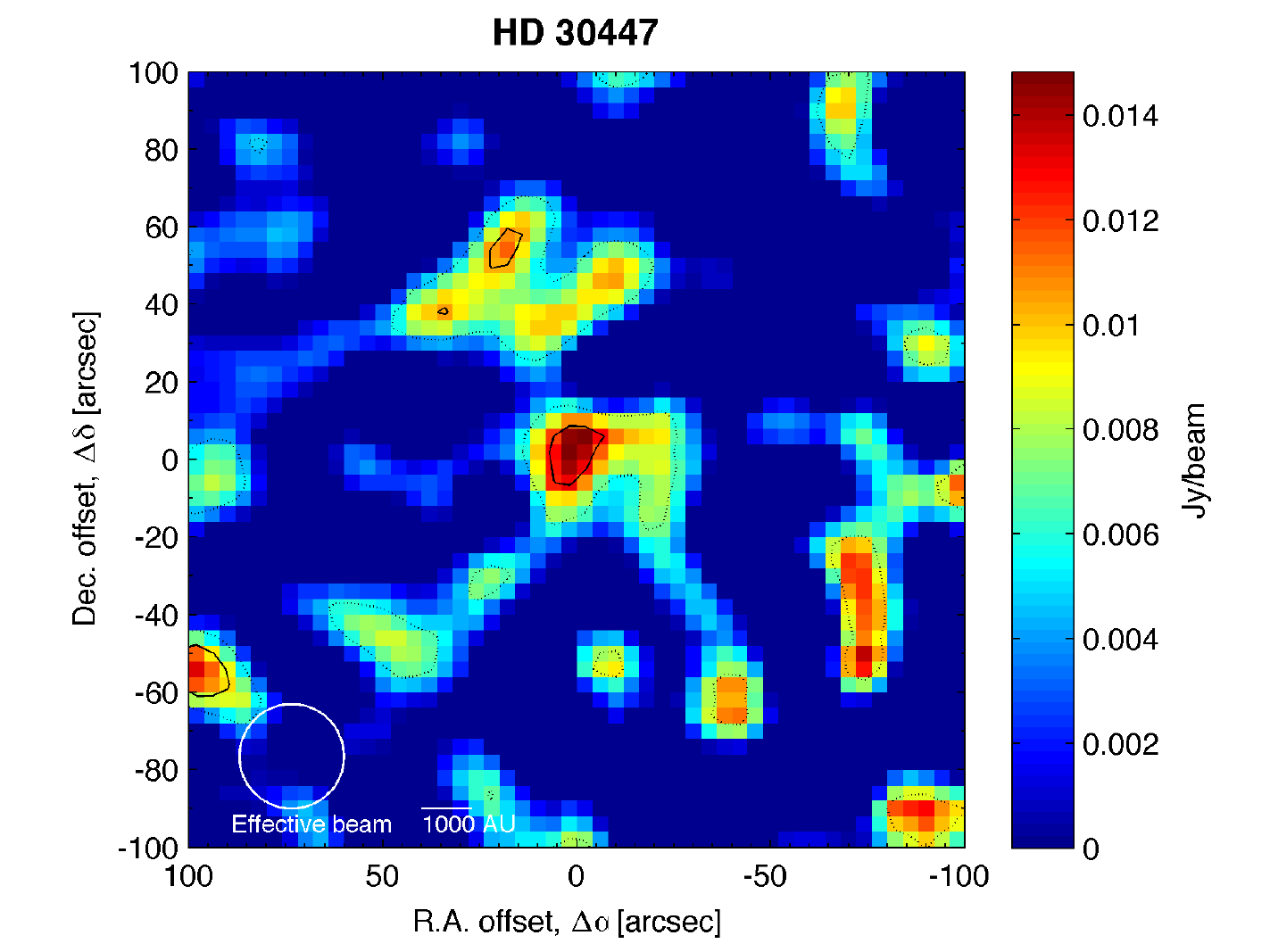}}
     \hspace{1mm}
     \subfigure[]{
          \label{fig:A1d}
          \includegraphics[width=.45\textwidth]{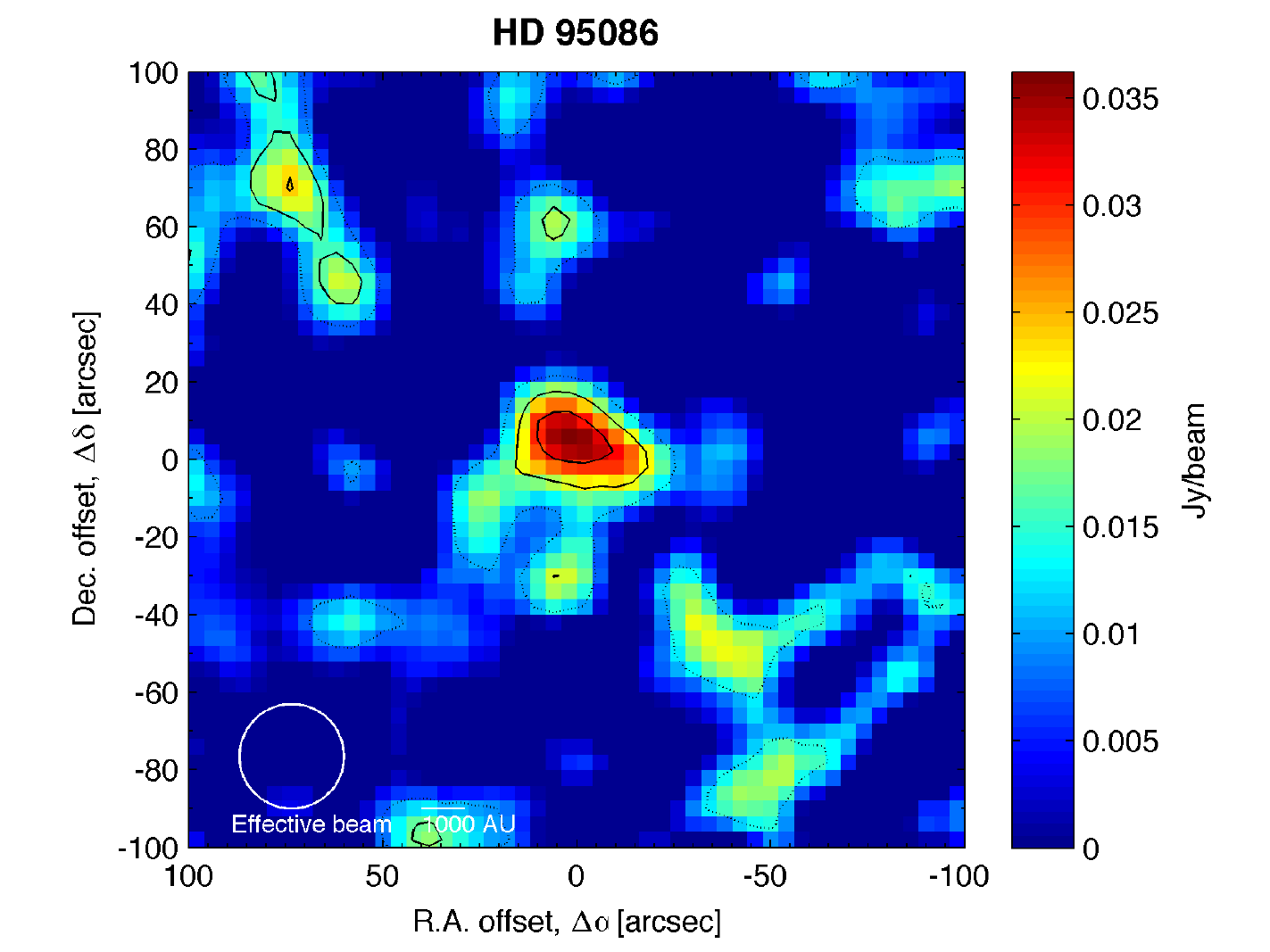}} \\
     \vspace{1mm}
     \subfigure[]{
          \label{fig:A1e}
          \includegraphics[width=.45\textwidth]{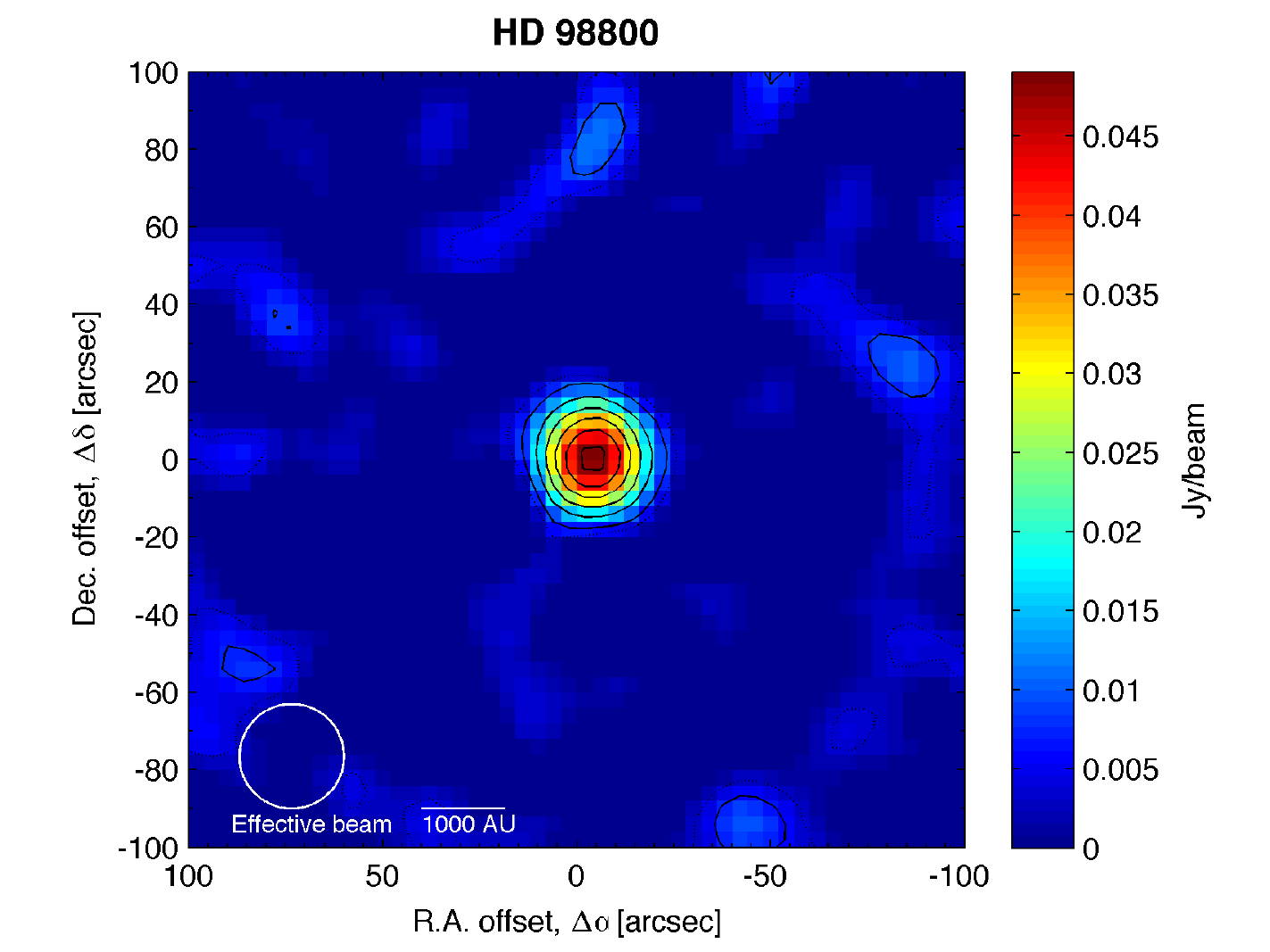}}
     \hspace{1mm}
     \subfigure[]{
          \label{fig:A1f}
          \includegraphics[width=.45\textwidth]{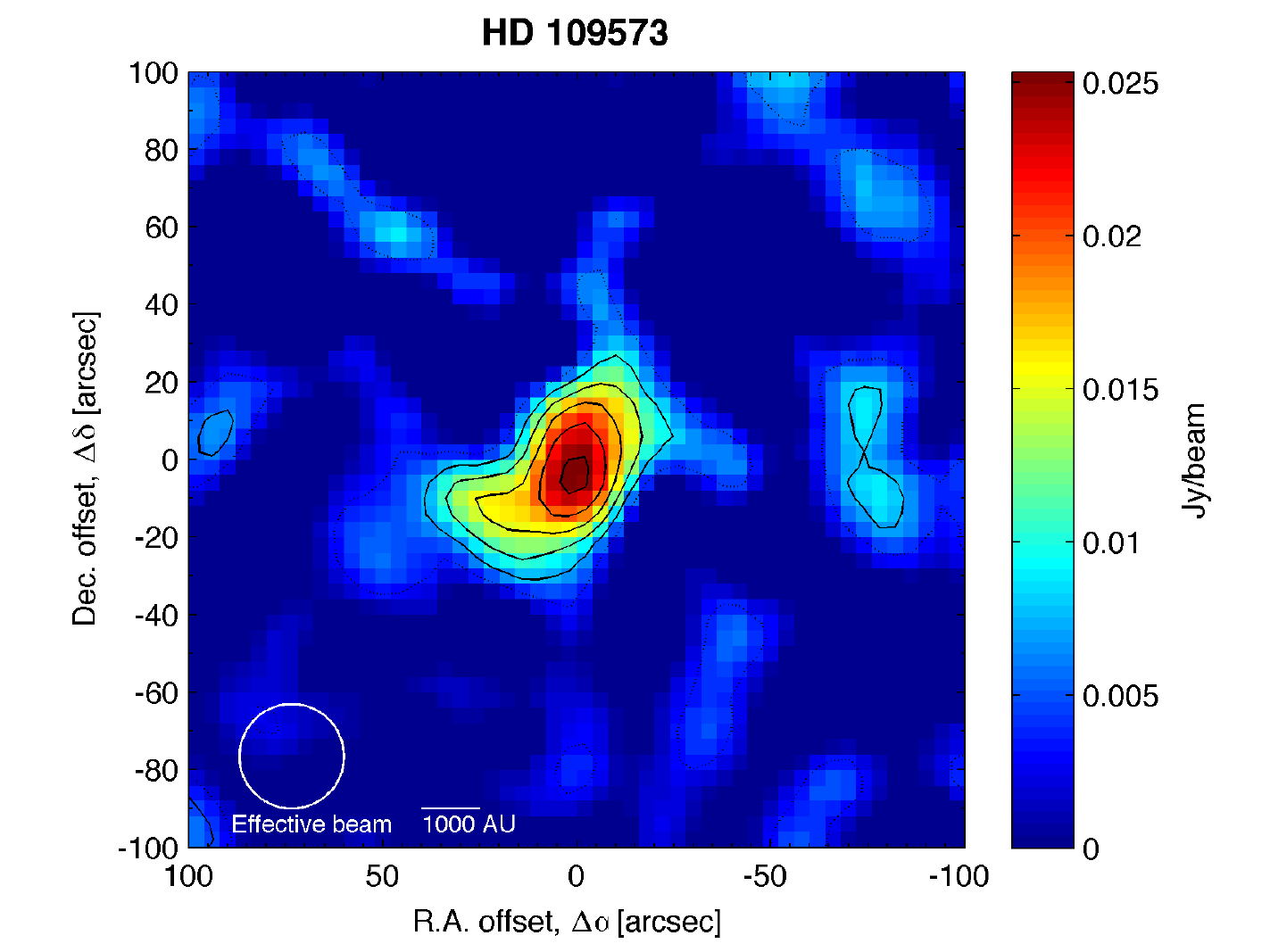}} \\
     \vspace{1mm}
\end{figure*}
\begin{figure*}
\centering
     \subfigure[]{
           \label{fig:A1g}
          \includegraphics[width=.45\textwidth]{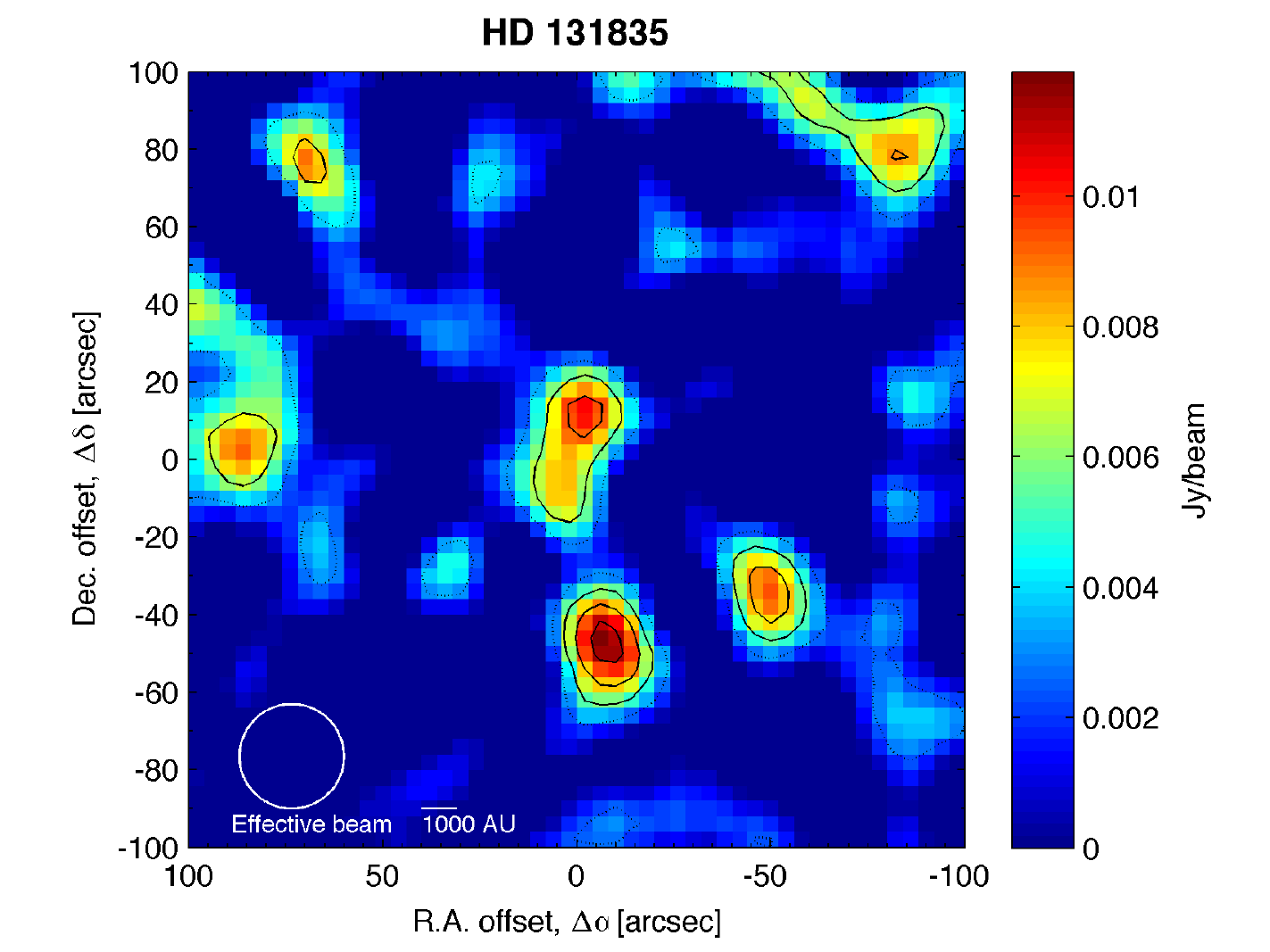}}
     \hspace{1mm}
     \subfigure[]{
          \label{fig:A1h}
          \includegraphics[width=.45\textwidth]{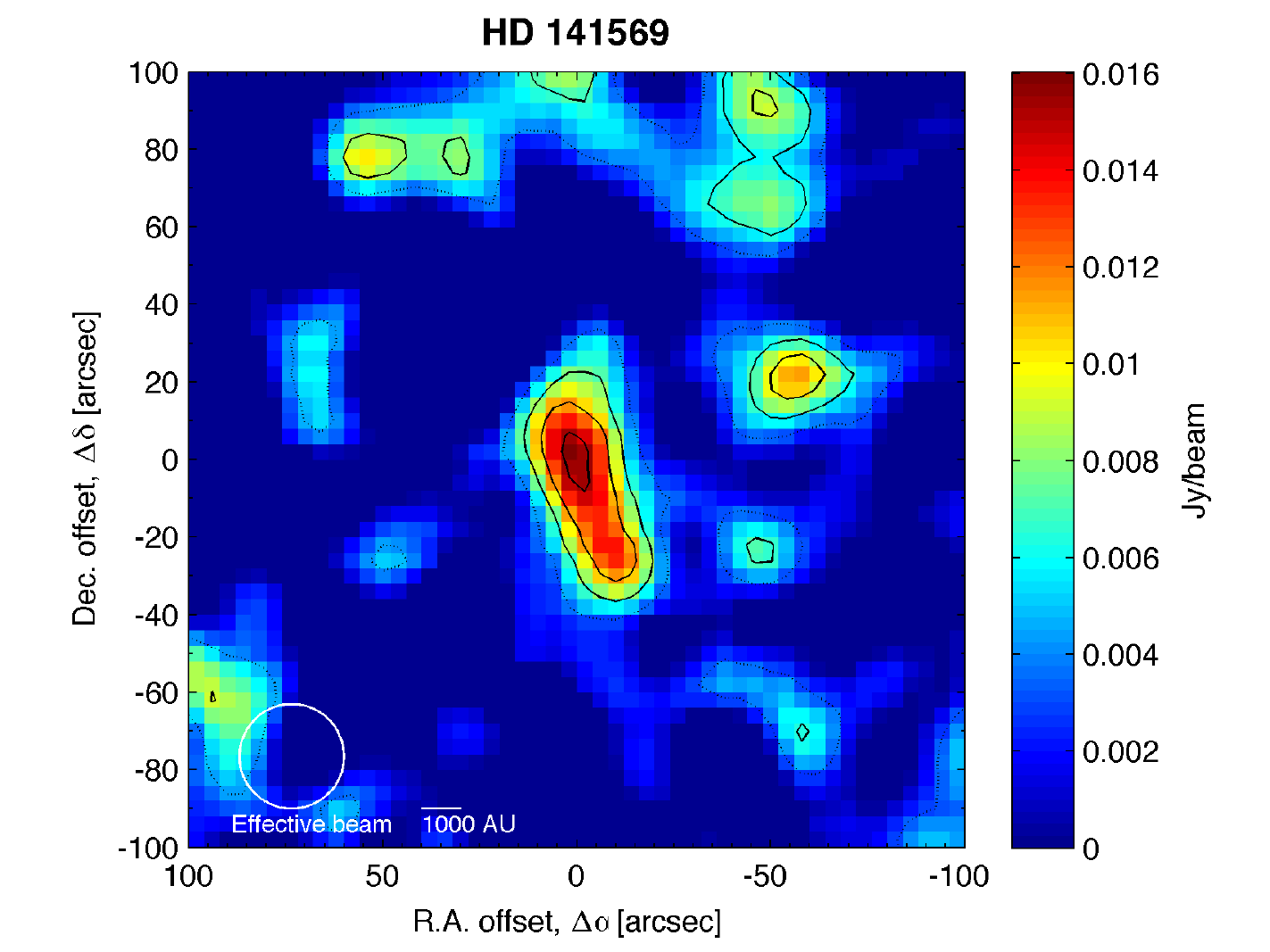}} \\
     \vspace{1mm}
     \subfigure[]{
          \label{fig:A1i}
          \includegraphics[width=.45\textwidth]{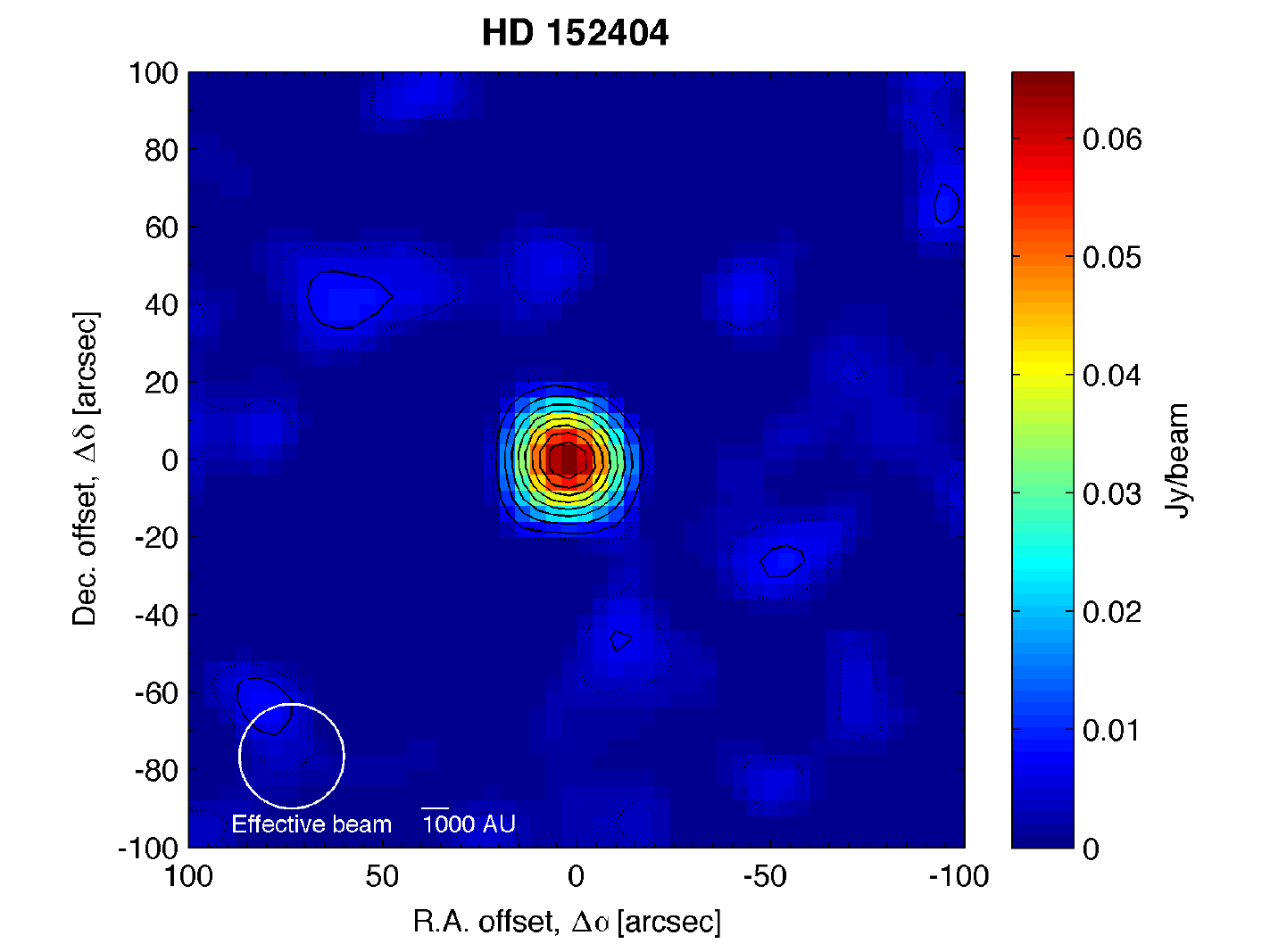}}
     \hspace{1mm}
     \subfigure[]{
          \label{fig:A1j}
          \includegraphics[width=.45\textwidth]{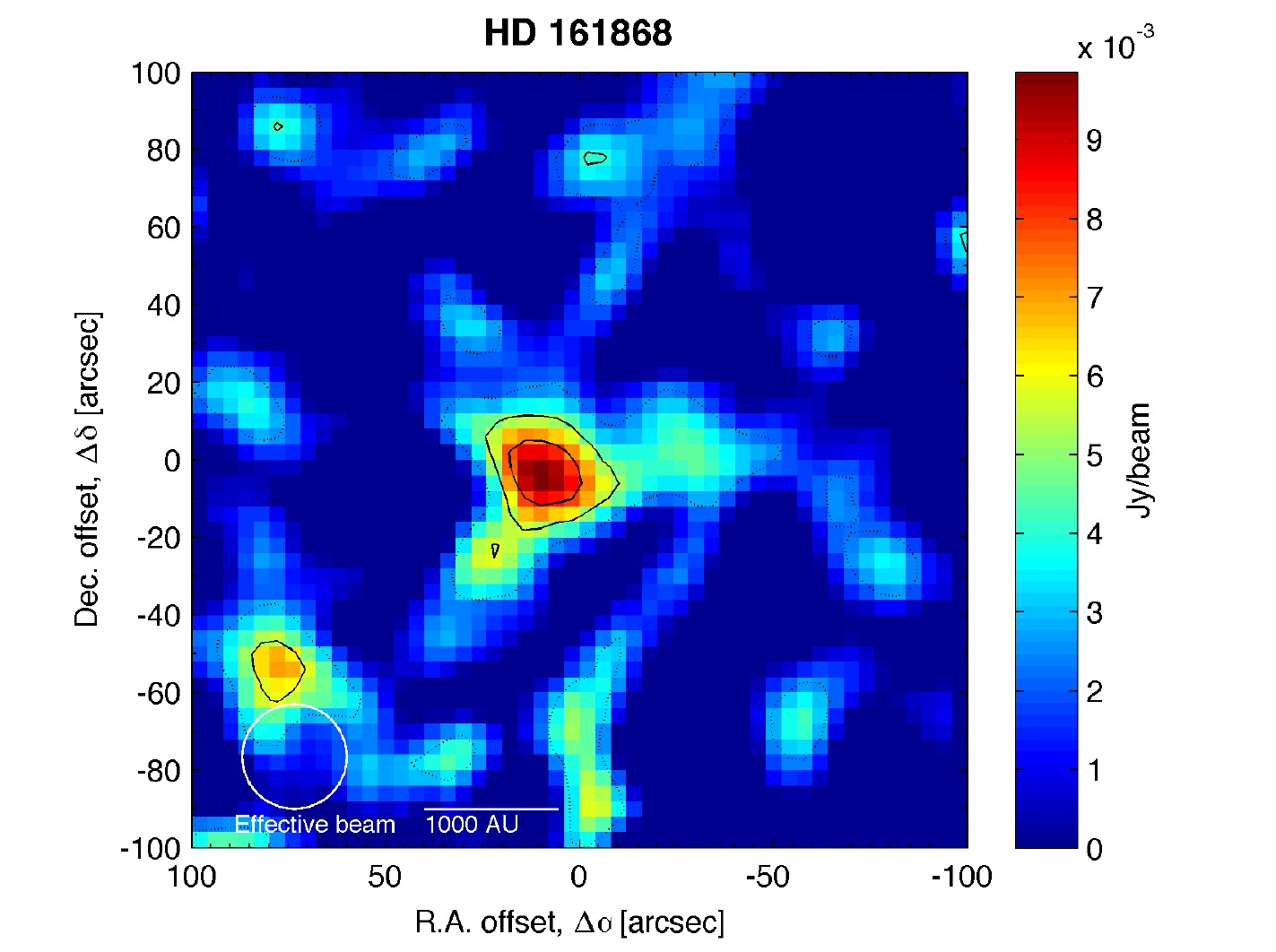}} \\
     \vspace{1mm}
     \subfigure[]{
           \label{fig:A1k}
          \includegraphics[width=.45\textwidth]{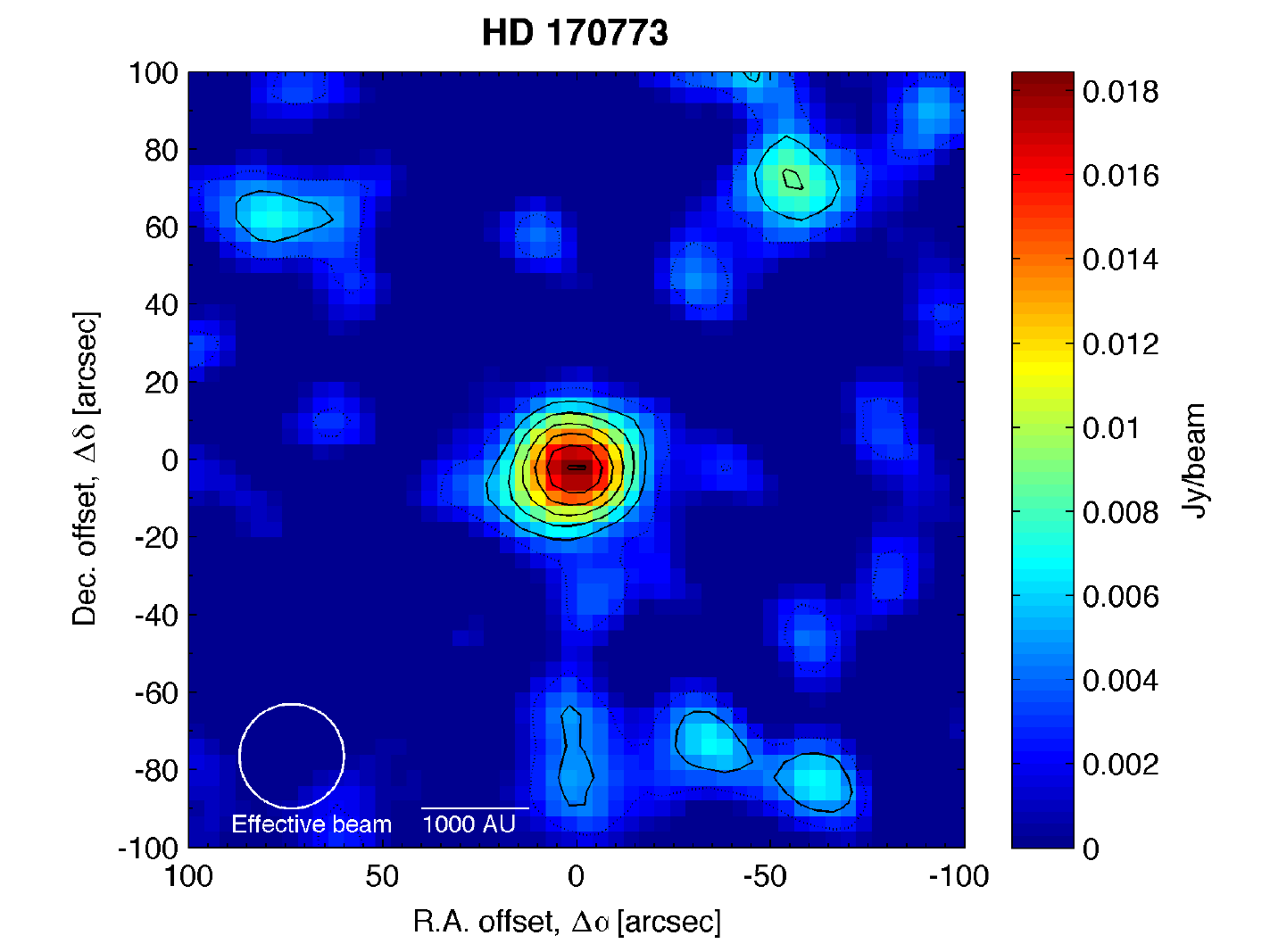}}
     \hspace{1mm}
     \subfigure[]{
          \label{fig:A1l}
          \includegraphics[width=.45\textwidth]{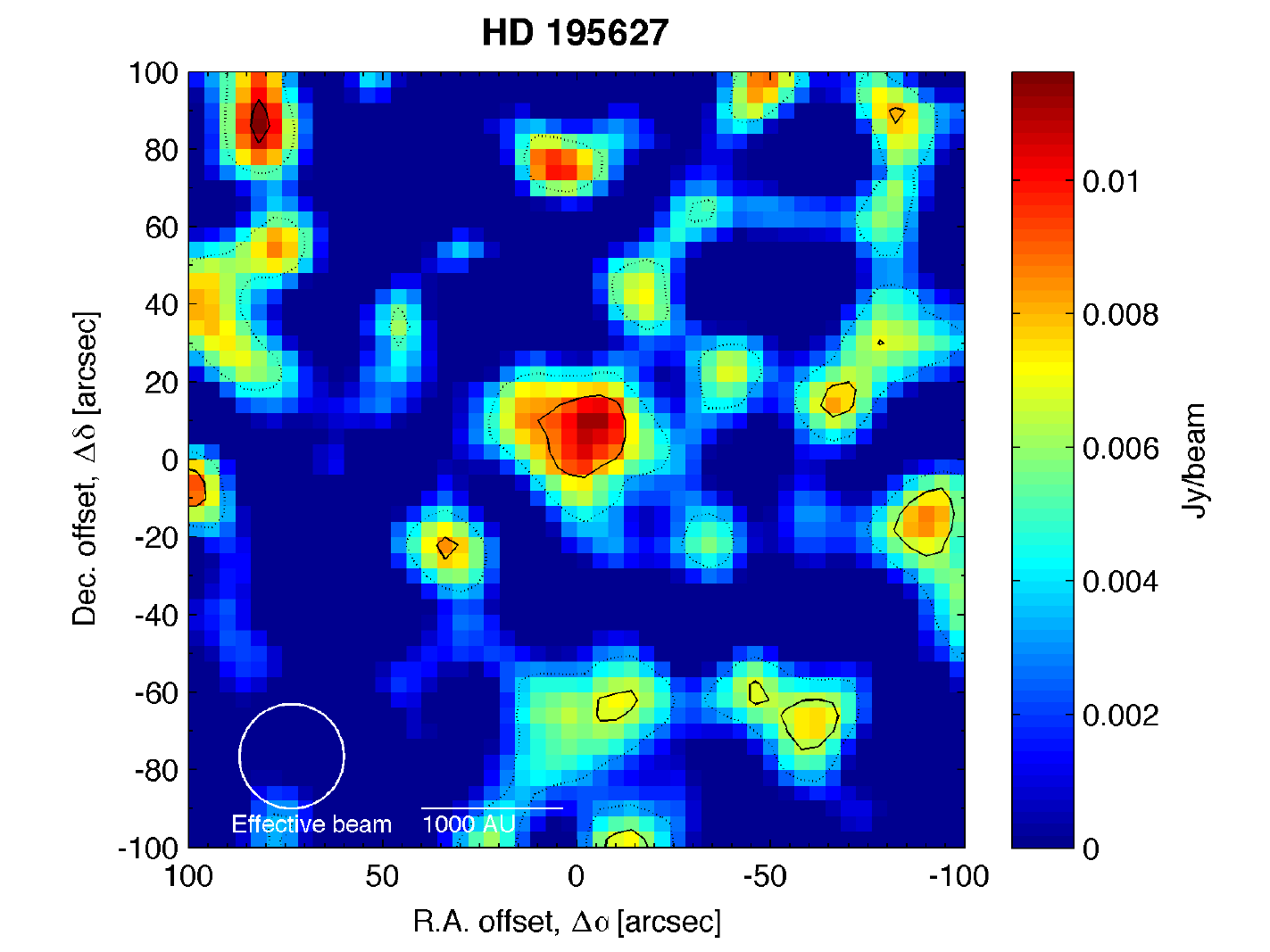}} \\
     \vspace{1mm}
\end{figure*}


\section{Discussion}\label{sec:dis}

\subsection{Frequency of Cold Dust Disks}
Similar to the results of \citet{Carpenter2005}, we find that about half of the stars previously observed to have far-IR fluxes in excess of stellar photospheric flux also have an excess of submm continuum emission. The five new submm disks discovered increase the number of known exo-Kuiper-Belt systems from 29 to 34 (see Appendix~\ref{sec:appenb}), counting the possible transitional disks\footnote{HD\,98800, HD\,141569, and HD\,152404 are, according to spectroscopic studies by e.g.\ \citet{Furlan2007}, \citet{Merin2004}, and \citet{Alencar2003}, respectively, pre-main-sequence stars and probably do not possess actual debris disks in the sense that they contains a large amount of gas, and might thus fall within the loosely defined transition disk category.} (HD\,98800, HD\,141569, HD\,152404), and with three additional sources (HD\,105, HD\,30447, and HD\,195627) awaiting confirmation. Although it is hard to estimate the total frequency of cold extended disks around main-sequence stars, it is likely to be at least the estimate of 10--15\% suggested from \emph{IRAS} \citep[e.g.][]{Backman1993,Rhee2007} and \emph{Spitzer} \citep{Hillenbrand2008a} detections of debris disks around main-sequence stars, and may be up to 25\% including disks too faint or cold to have been detected with current instrument sensitivities \citep{Matthews2007}.

An interesting result is that all four multiple systems in our sample, HD\,98800, HD\,109573, HD\,141569, and HD\,152404, were detected at 870\,$\mu$m. Multiple stars have previously been deliberately left out from most debris disk surveys under the assumption that the formation of planetesimals in such systems would be inhibited. However, the \emph{Spitzer} MIPS observations presented by \citet{Trilling2007} show that the incidence of debris disks around main-sequence A3--F8 wide ($>$\,50\,AU) binaries is comparable to that around single stars and is even marginally higher for tight ($<$3\,AU separation) binaries. Our results confirm these trends, since the four multiple star systems in our sample are three wide binaries and one spectroscopic tight binary (HD152404). These results should however be taken with caution because of the morphological states of these four multiple stars systems. The main one is that all four systems are young ($\sim$5--15\,Myr old) and thus should naturally have a higher probability of possessing massive debris disks (see discussion in Section\,\ref{sec:tempevo}). This bias towards young stars in our sample of multiple star systems make it difficult to infer trends related to the binarity. 

The fraction of detected debris disks in this first round of observations (45\%) is considerably higher than that of previous submm surveys, e.g.\ that of nearby bright stars \citep[9\%,][]{Holmes2003}, {$\beta$} Pictoris Moving Group \citep[17\%,][]{Nilsson2009,Liu2004}, FEPS nearby stars \citep[3\%,][]{Carpenter2005}, nearby G stars \citep[15\%,][]{Greaves2005}, and M dwarfs \citep[6\%,][]{Lestrade2006}. This cannot only be explained by higher instrument sensitivity but is most likely due to a selection effect, having chosen an initial sample of stars with high far-IR excesses, and we are not expecting such a high detection rate from our ongoing Large Programme observations.

As anticipated with RMS noise levels as low as 2\,mJy/beam we find a number of significant flux density peaks originating from background submm galaxies. Based on results from previous deep galaxy surveys at mm and submm wavelengths \citep[e.g.][]{Weiss2009,Bertoldi2007,Ivison2007,Clements2004} we would expect about 4--6 extragalactic sources to be detected (at $>$3$\sigma$ levels) in the LABOCA field-of-view with this sensitivity. This is consistent with the average number of four 3$\sigma$ peaks within a 4$\arcmin$ radius around the central position of the source that we observe in our final maps. We can estimate the probability of a finding a random background source within an angular distance $r$ of a given position from the expression $P(<r)=1-\exp({\pi}{\rho}r^{2})$ \citep{Condon1998}, where $\rho$ is the source density per square degree. With a 27$\arcsec$ effective beam we calculate a roughly 2\% likelihood of a chance alignment. Although the risk of a background galaxy falling within the central beam is so small, the probability that it does must be considered, as it could produce a false disk detection or a seemingly extended disk. With the commissioning of the \emph{Atacama Large Millimetre/Submillimetre Array} (ALMA) now on its way, unprecedented angular resolution \citep[$\gtrsim$$0\farcs01$,][]{Peck2008} at submm and mm wavelengths (from 350\,$\mu$m to 10\,mm) will soon permit such ambiguities to be resolved.

\begin{figure*}
     \centering
     \subfigure[]{
          \label{fig:B1a}
	\includegraphics[width=.45\textwidth]{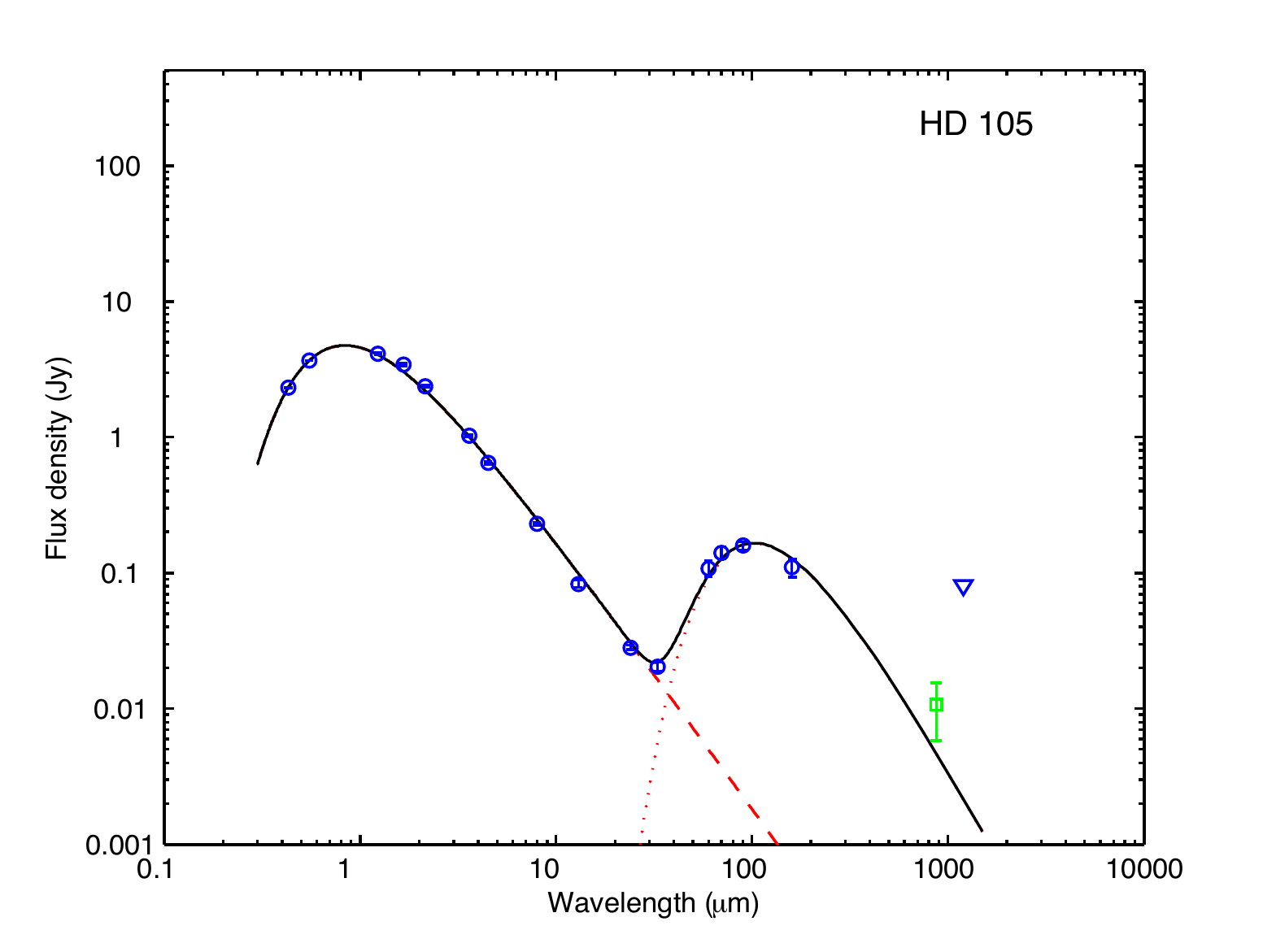}}
     \hspace{1mm}
     \subfigure[]{
          \label{fig:B1b}
    	\includegraphics[width=.45\textwidth]{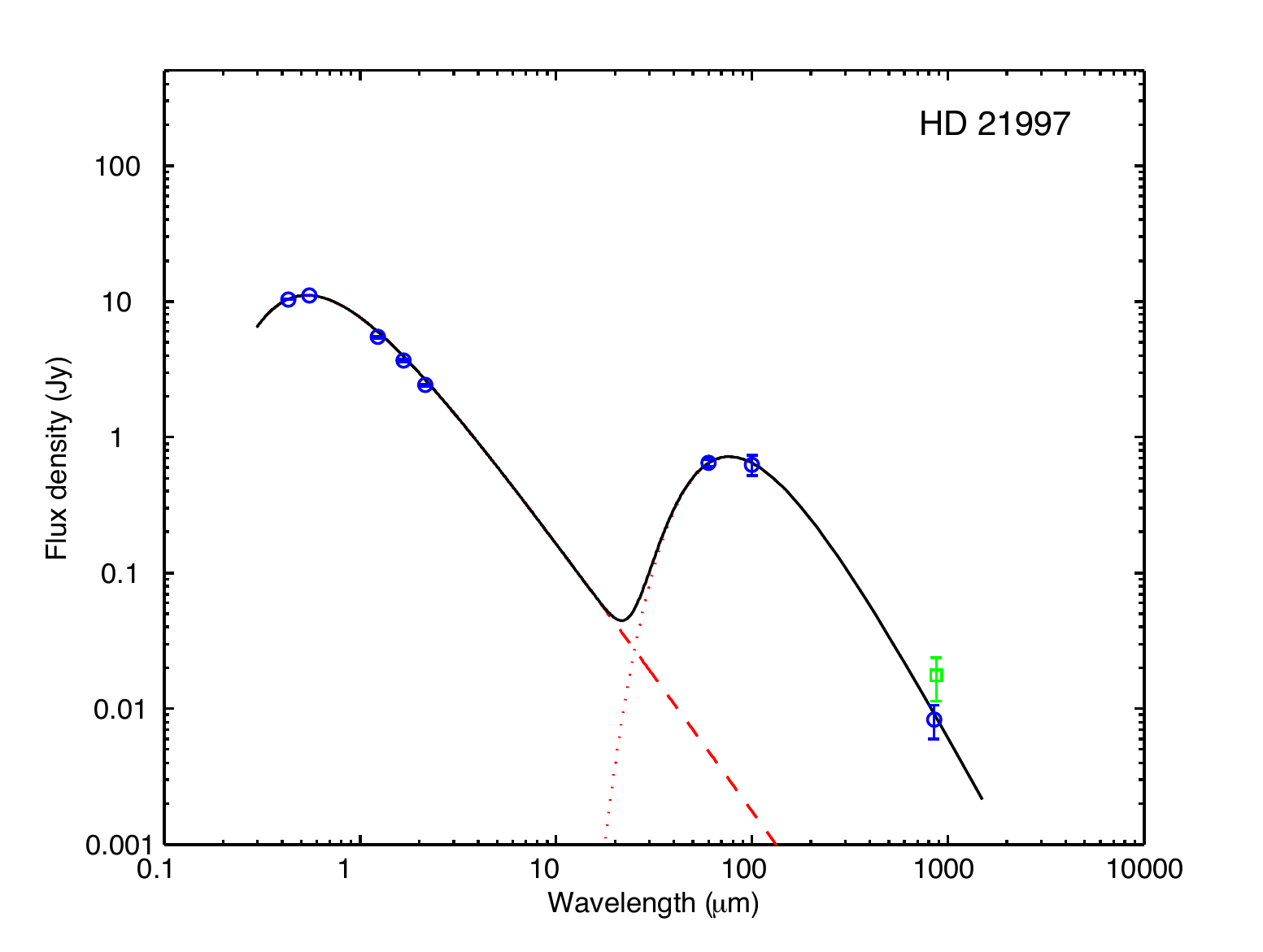}} \\
     \vspace{1mm}
     \subfigure[]{
           \label{fig:B1c}
	\includegraphics[width=.45\textwidth]{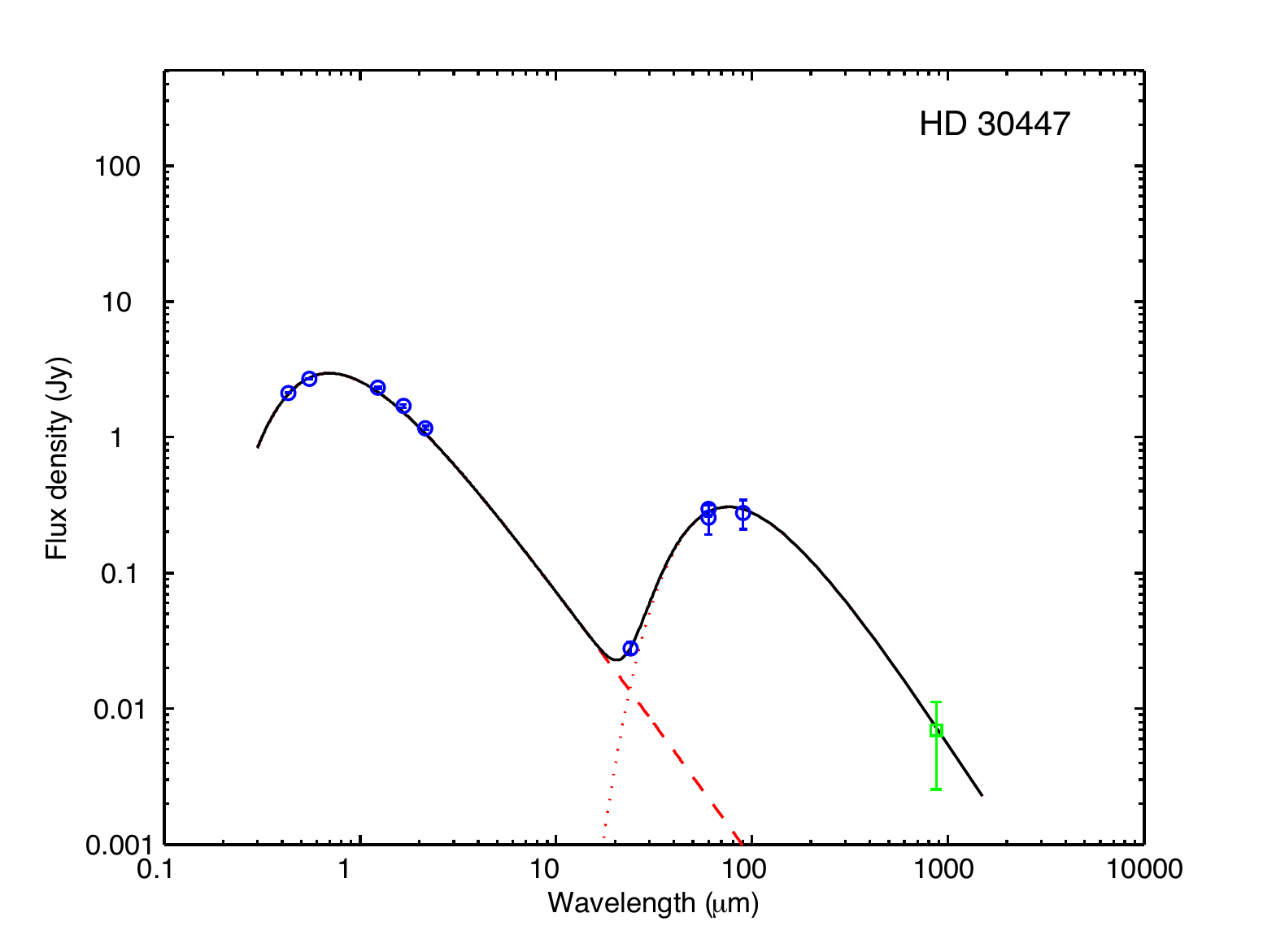}}
     \hspace{1mm}
     \subfigure[]{
          \label{fig:B1d}
	\includegraphics[width=.45\textwidth]{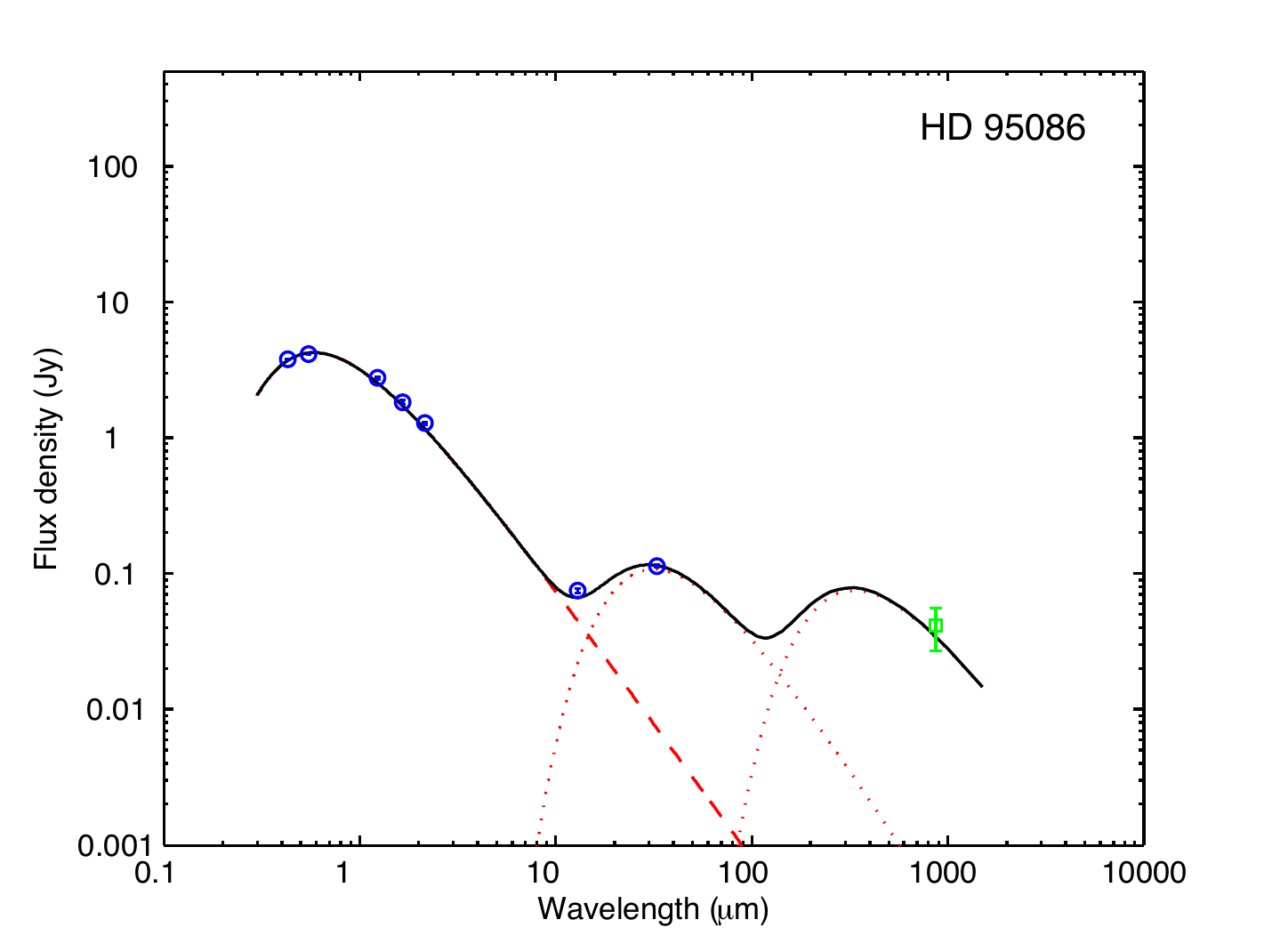}} \\
     \vspace{1mm}
     \subfigure[]{
          \label{fig:B1e}
	\includegraphics[width=.45\textwidth]{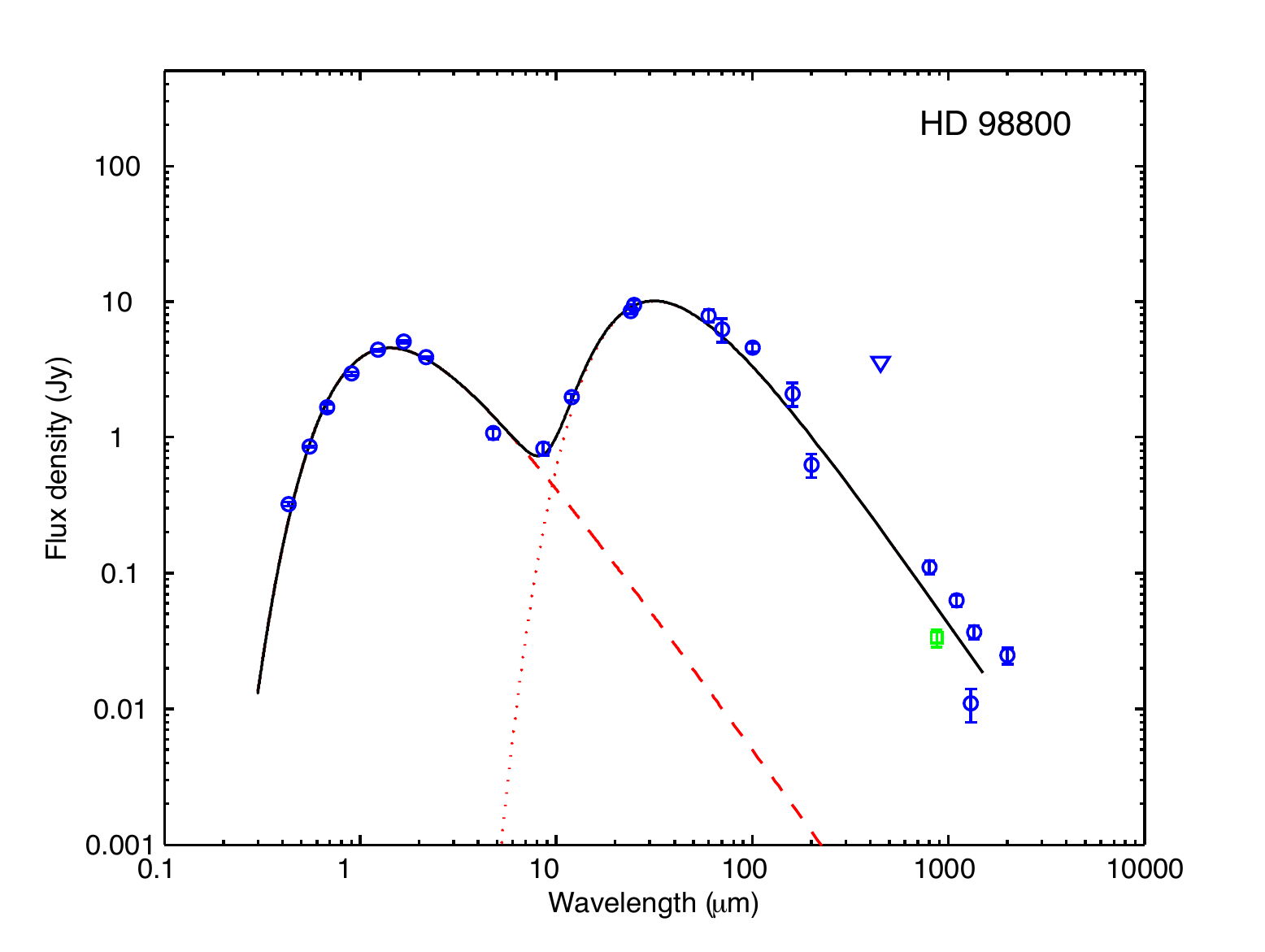}}
     \hspace{1mm}
     \subfigure[]{
          \label{fig:B1f}
	\includegraphics[width=.45\textwidth]{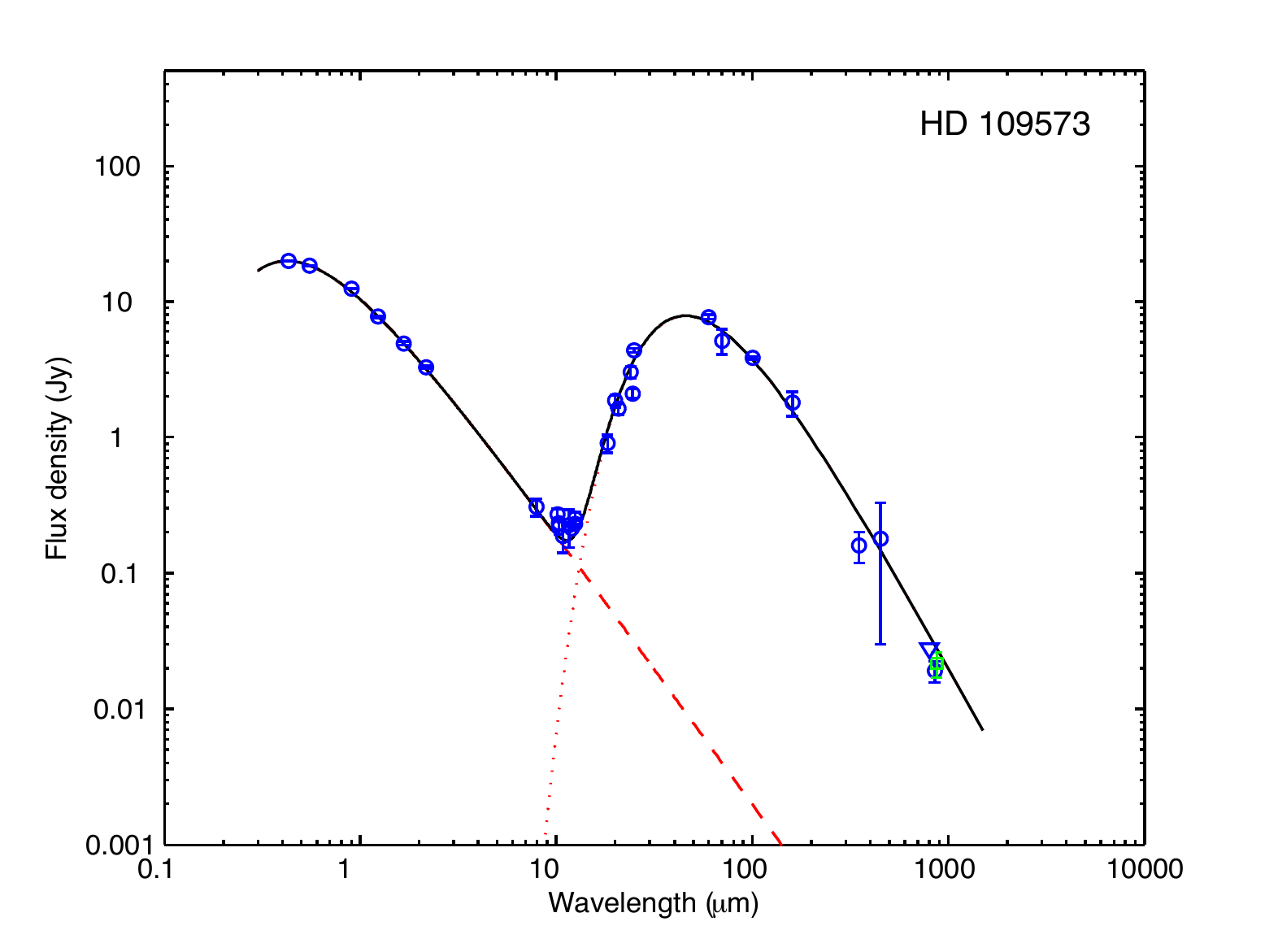}} \\
     \vspace{1mm}
\end{figure*}
\begin{figure*}
\centering
     \subfigure[]{
           \label{fig:B1g}
	\includegraphics[width=.45\textwidth]{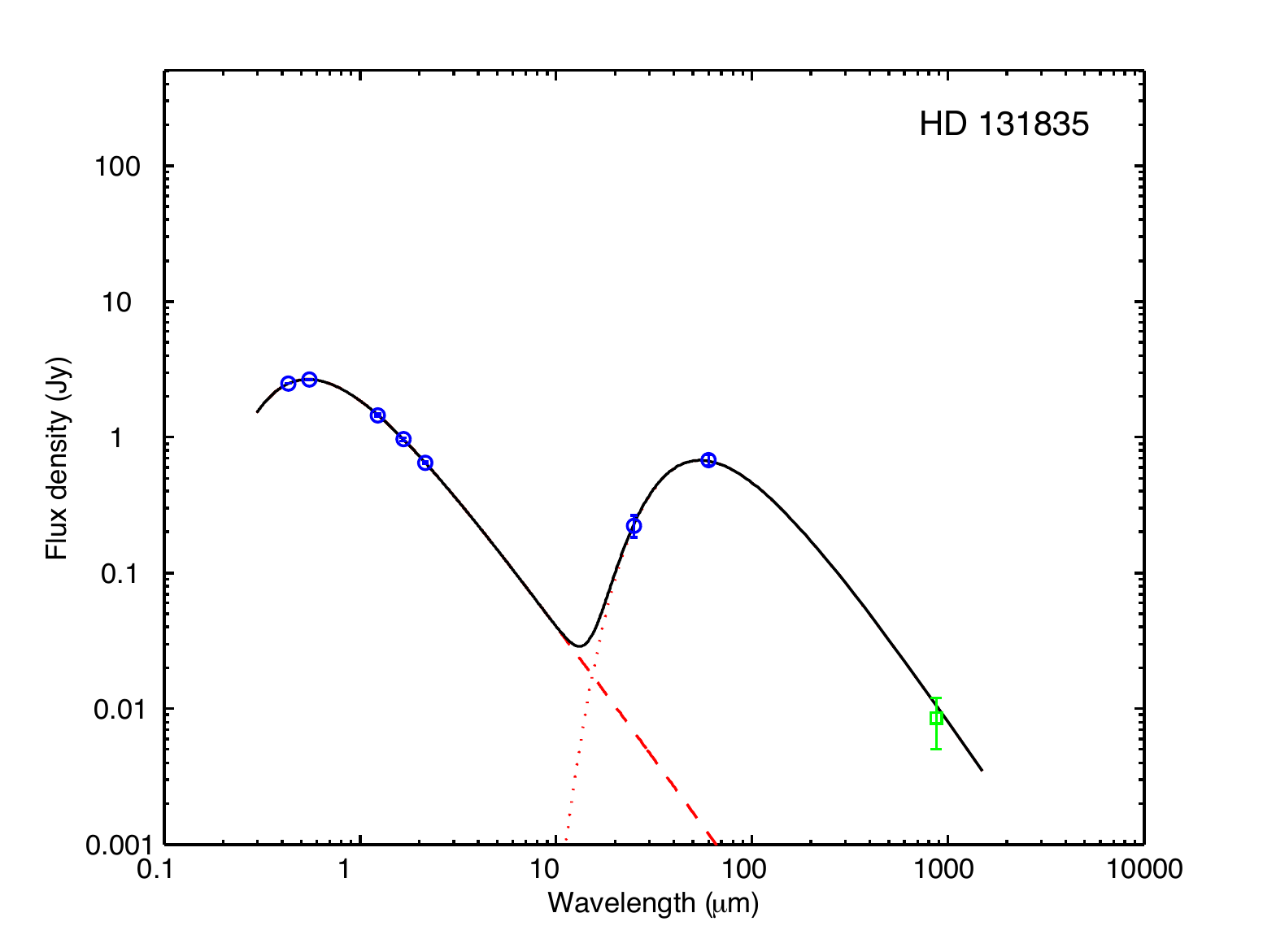}}
     \hspace{1mm}
     \subfigure[]{
          \label{fig:B1h}
	\includegraphics[width=.45\textwidth]{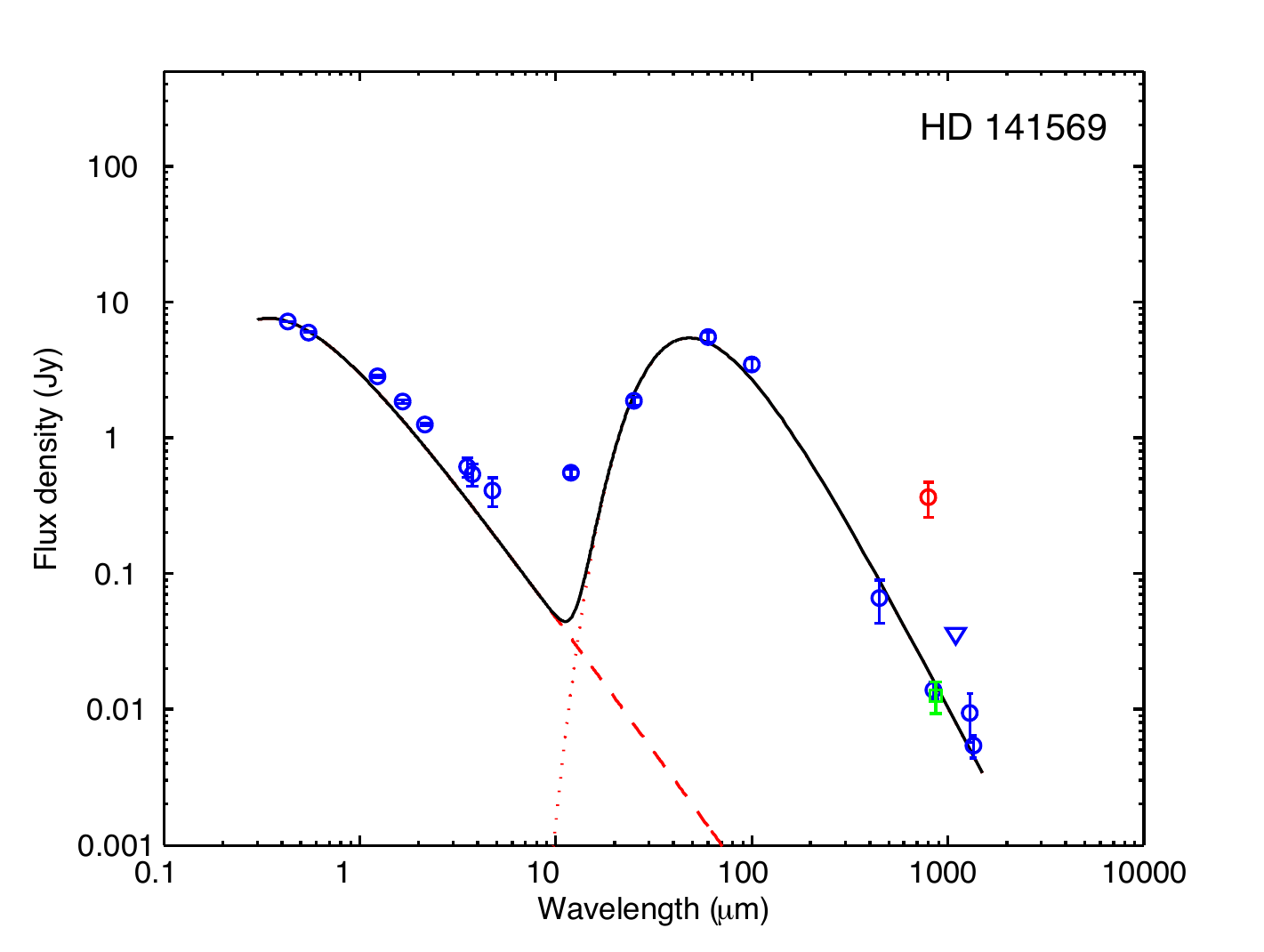}} \\
     \vspace{1mm}
     \subfigure[]{
          \label{fig:B1i}
	\includegraphics[width=.45\textwidth]{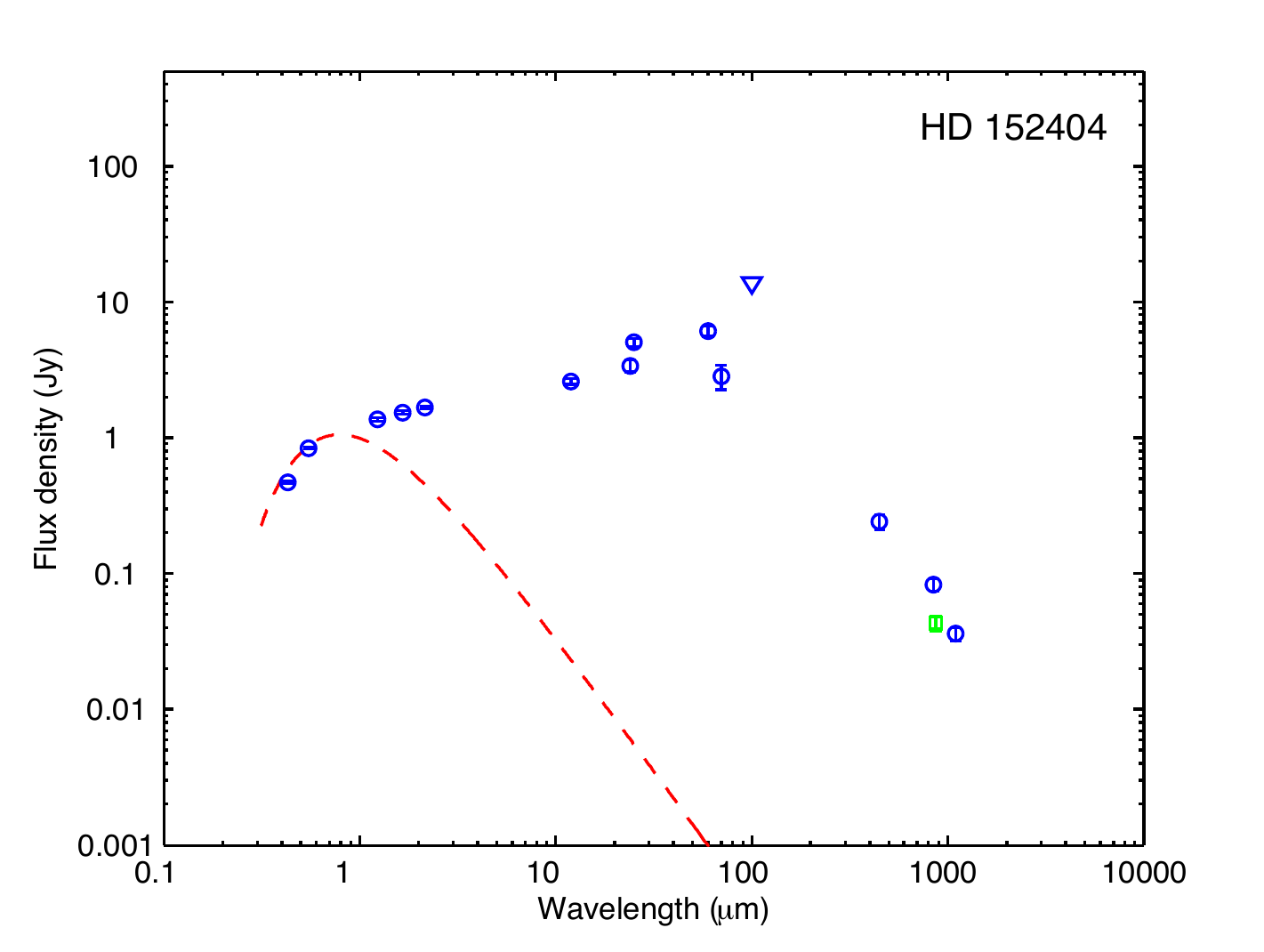}}
     \hspace{1mm}
     \subfigure[]{
          \label{fig:B1j}
	\includegraphics[width=.45\textwidth]{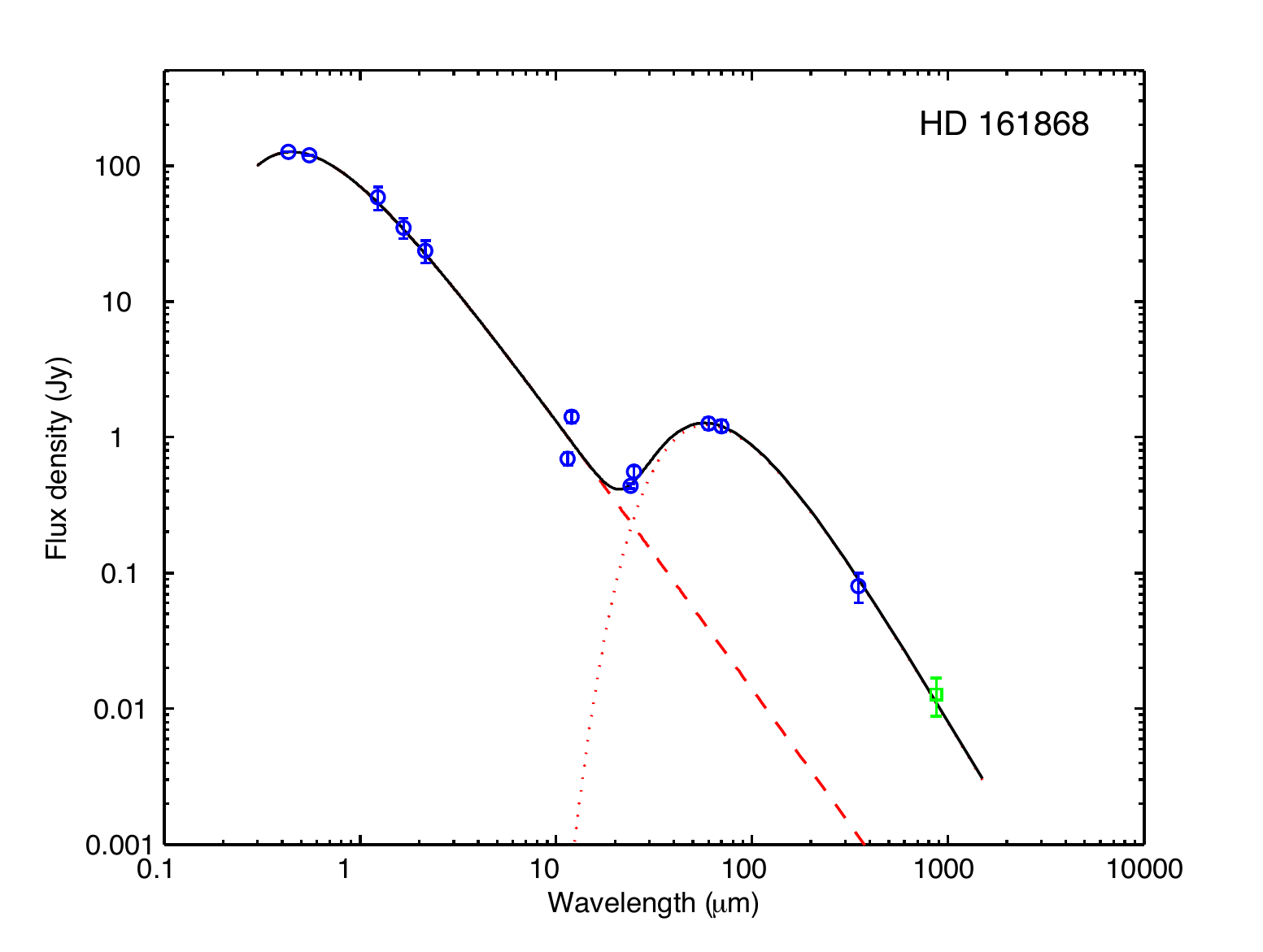}} \\
     \vspace{1mm}
     \subfigure[]{
           \label{fig:B1k}
	\includegraphics[width=.45\textwidth]{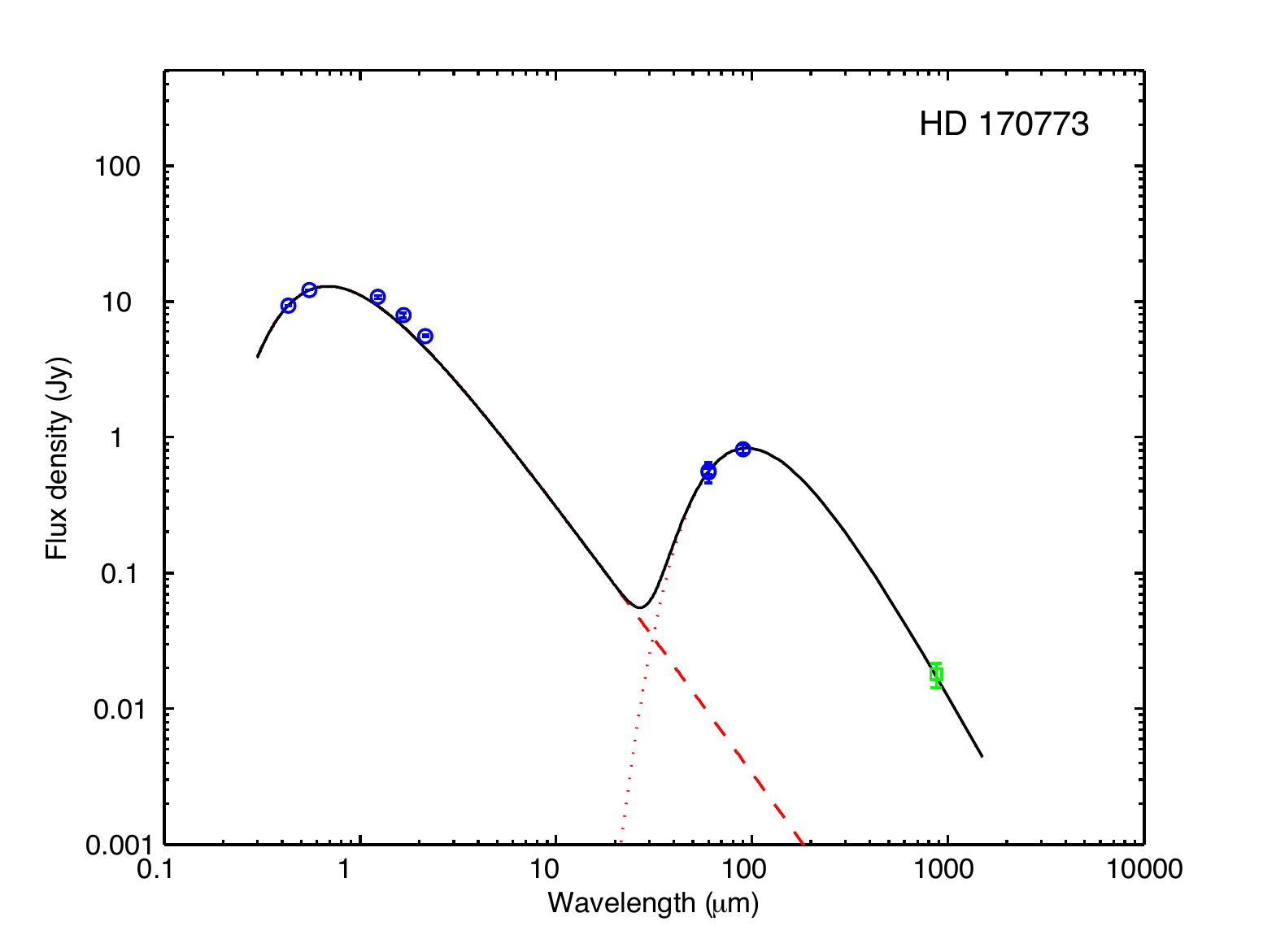}}
     \hspace{1mm}
     \subfigure[]{
          \label{fig:B1l}
	\includegraphics[width=.45\textwidth]{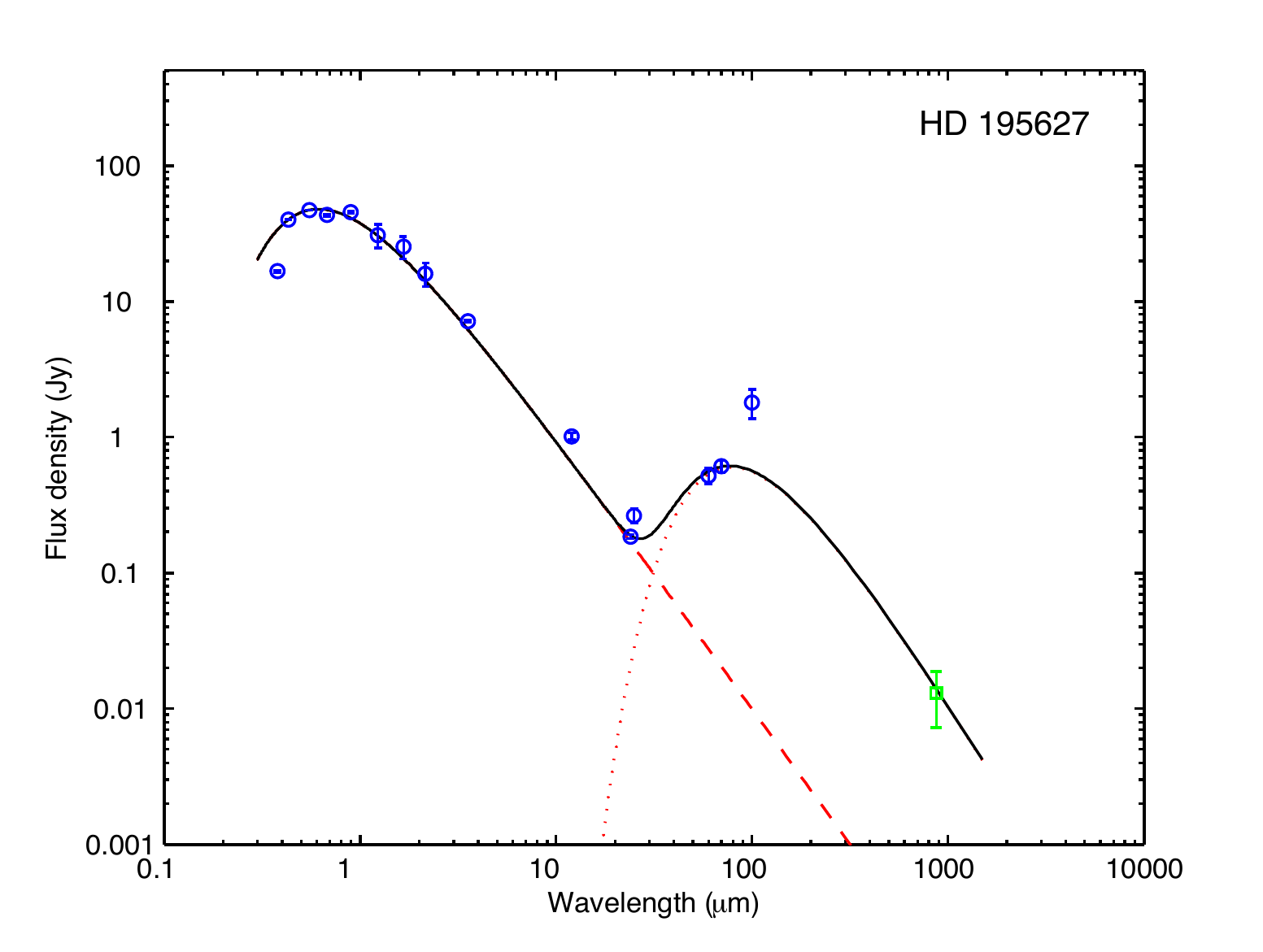}} \\
\end{figure*} 
\begin{figure*}
     \centering
     \subfigure[]{
          \label{fig:B1m}
	\includegraphics[width=.45\textwidth]{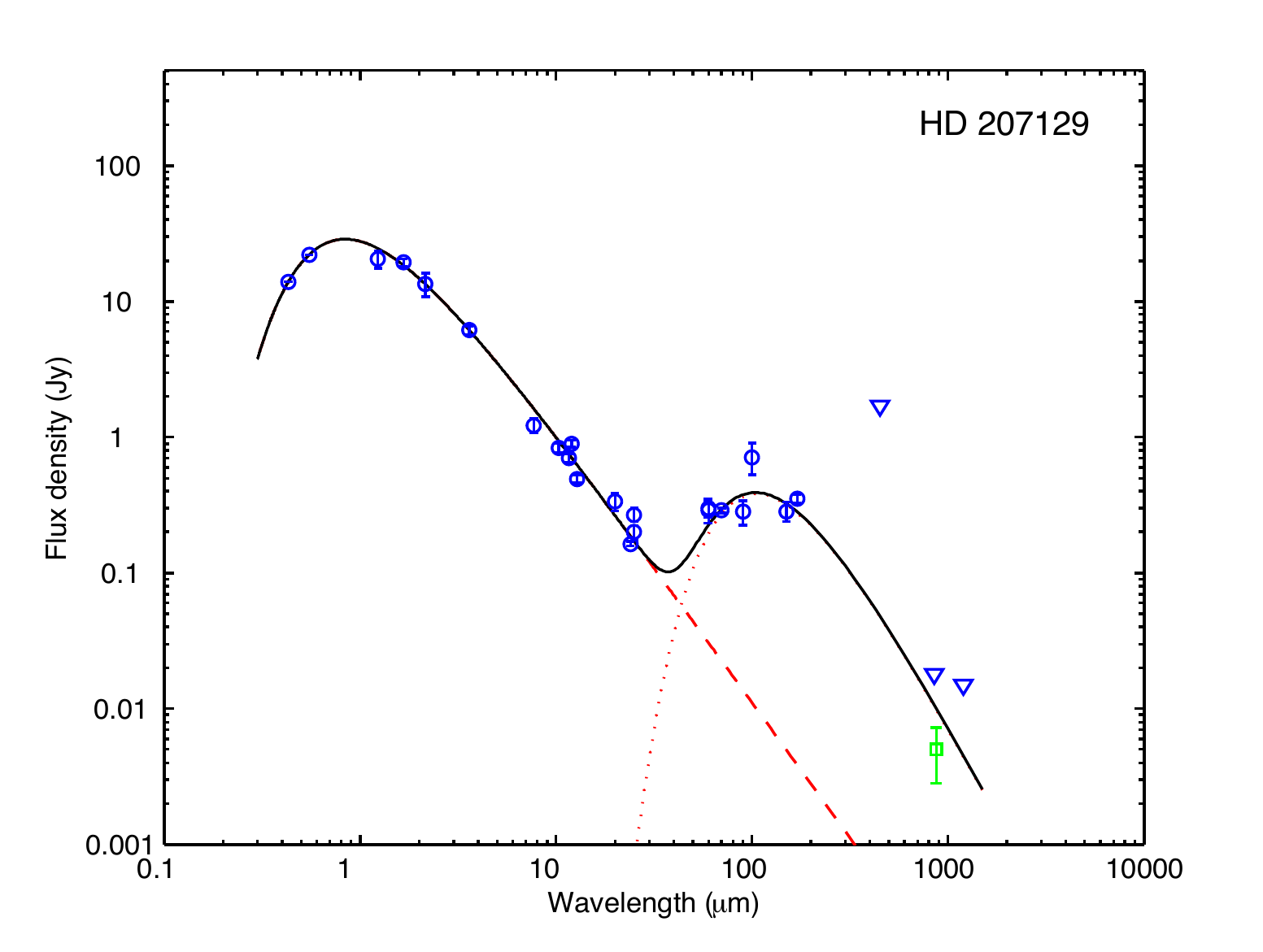}} \\
     \caption{The total modelled SEDs of detected objects are shown as a solid black line, with the stellar and disk components represented by red dashed and dotted lines, respectively. Our obtained 870\,$\mu$m integrated flux density is given by a green square, while blue circular markers (and triangles symbolizing upper limits) are data from the literature. Fits which yield exceptionally large ${\chi}^{2}$ values and other special cases are discussed in Section\,\ref{sec:sed}. The 800\,$\mu$m flux measured for HD\,141569 (red marker) at CSO \citep{Walker1995} has been left out of the fit since it is probably incorrect. 
\textsc{References}: HD\,105 \citep{Moor2006,Hillenbrand2008a,Carpenter2005,Cutri2003}; HD\,21997 \citep{Moor2006,Williams2006,Cutri2003}; HD\,30447 \citep{Moor2006,Cutri2003}; HD\,95086 \citep{Chen2006,Cutri2003}; HD\,98800 \citep{Low2005,Schutz2005,Mannings1998,Cutri2003,Walker2000,Gregorio-Hetem1992,Garcia-Lario1990,Sylvester1996,Cieza2008,Sylvester2001}; HD\,109573 \citep{Moor2006,Sheret2004,Barrado2006,Cutri2003,Low2005,Jura1995,Telesco2000,Jura1993,Fajardo-Acosta1998,Wahhaj2005}; HD\,131835 \citep{Moor2006,Cutri2003}; HD\,141569 \citep{Sylvester2001,Sylvester1996,Beichman1988,Walker1995,Sheret2004}; HD\,152404 \citep{Chen2005b,Jensen1996,Acke2004,Cutri2003}; HD\,161868 \citep{Su2008,Fajardo-Acosta1999,Jaschek1991,Cutri2003}; HD\,170773 \citep{Moor2006,Cutri2003}; HD\,195627 \citep{Rebull2008,Sylvester2000,Mannings1998,Cutri2003}; HD\,207129 \citep{Mannings1998,Trilling2008,Schutz2005,Cutri2003,Mamajek2004,Laureijs2002,JourdaindeMuizon1999,Tanner2009}.}
     \label{fig:B1all}
\end{figure*}

\subsection{Spectral Energy Distributions}\label{sec:sed}
By combining the integrated submm flux of our detected objects with photometry from optical and infrared measurements found in the literature, we made fits to the SED in order to find the dust temperature and radial extent of the disk. The star is modelled as a simple scaled blackbody, while the disk is approximated with a modified scaled blackbody function, as
\begin{equation}
\begin{split}\label{eqn:1}
F_{\mathrm{disk}}(\nu)& =\frac{2h{\nu}^{3}}{(\exp\left(\frac{h\nu}{kT_{\mathrm{dust}}}\right)-1)c^{2}}\left(\frac{\nu}{\nu_{0}}\right)^{\beta}{\kappa}_{0}\Omega \\
& =\frac{{\nu}^{3+\beta}}{(\exp\left(\frac{h\nu}{kT_{\mathrm{dust}}}\right)-1)}C,
\end{split}
\end{equation}
where $h$ is Planck's constant, $k$ is the Boltzmann constant, and $c$ is the speed of light, with the radiating surface area and distance incorporated in the scale factor $C$. Here we assume that the opacity index varies with the frequency as a power law, i.e.\ ${\kappa_{\nu}}={\kappa_{0}}({\nu}/{\nu_{0}})^{\beta}$, and that the grains dominating the flux in the IR and submm all have the same temperature. The latter simplifying assumption is applied since we do not have a spatially resolved source at submm wavelengths, although it would be reasonable to believe that colder dust located at Kuiper-belt distances would be responsible for most of the submm emission, while warmer dust located at asteroid-belt distances would dominate in the far-IR. Best-fit parameters were found by minimizing the reduced $\chi^{2}$, i.e. the sum of squared errors weighted with the accuracy of the individual photometry measurements, divided by the number of degrees of freedom. Resulting SED fits are presented in Fig.\,\ref{fig:B1all}, while values of the dust temperature, $T_{\mathrm{dust}}$, and the power law exponent of the opacity law, $\beta$, for the 10 detected (and 3 marginally detected) objects are listed in Table\,\ref{table:2}.

Although only the best-fit $\beta$ parameter is quoted (without errors), we note that in almost all fits a value between 0.2 and 0.7 gives roughly the same $\chi^{2}$. The difficulty in determining $\beta$ is due to relatively large errors in measured submm (and longer wavelength) data, in addition to the possible existence of colder dust that would have been better modelled with a second dust component, but now instead shifts $\beta$ in a single component fit towards zero. One example is HD\,95086 which clearly cannot be modelled with a single component dust disk. For this star we added a second disk with a lower temperature limited by the maximum extent of the unresolved disk ($\lesssim$1500\,AU) and upper limit constrained by the two IR points at 13 and 33\,$\mu$m \citep{Chen2006}. The fit shown in Fig.\,\ref{fig:B1e} is for a lower limit 15\,K disk. Overall, the derived $\beta$ values for our sources indicate significantly larger (and perhaps amorphous and/or fractal) dust grains \citep{Miyake1993,Mannings1994,Pollack1994} in circumstellar debris disks compared to the mostly unprocessed grains in the interstellar medium (ISM), which have ${\beta}{\sim}2$ \citep{Hildebrand1983,Li2001}. This is in agreement with previous results by e.g.\ \citet{Najita2005}, \citet{Nilsson2009}, and \citet{Roccatagliata2009} who found ${\beta}$=0--1 for various debris disks.

In three cases the SED modelling does not yield a good fit. For the quadruple system HD\,98800 (Fig.\,\ref{fig:B1e}) previous submm and mm photometry has produced somewhat differing results, viz.\ the 1350\,$\mu$m flux from \citet{Sylvester2001} is significantly higher than the 1300\,$\mu$m flux from \citet{Sylvester1996}, with our 870\,$\mu$m measurement more consistent with the latter for a typical Rayleigh-Jeans slope. It can not be excluded that processes related to the orbital phase of the tight (1\,AU) binary HD\,98800B cause this seemingly bimodal flux variation of its circumbinary disk. Photometric variability has been found at shorter wavelengths \citep{Soderblom1998}. In any case, this unusual system of two eccentric binaries orbiting each other with highly inclined orbital planes \citep[see e.g.][and references therein]{Verrier2008} should have a complicated dynamical effect on the debris (or transitional) disk around the B pair, e.g.\ the high IR-excess has been proposed to originate from a puffed up outer dust ring which is gravitationally perturbed by the HD\,98800A binary at perihelion (50\,AU) passage \citep{Furlan2007}.

For HD\,141569 the mid-IR flux is difficult to fit consistently with longer wavelength data in a simple disk model, however our results are very similar to those of \citet{Sheret2004}.

The photometry data of HD\,152404 (AK\,Sco) can clearly not be fitted with a simple modified blackbody emitting disk. In Fig.\,\ref{fig:B1i} we have inserted a dashed line representing the SED of the binary consisting of two F5V stars, showing that excess emission is present already at $\lambda$$\lesssim$1\,$\mu$m. This is indicative of hot dust in the system's young circumbinary disk, and it is probably still in its protoplanetary or transitional disk stage. 

Although more detailed modelling of the SED could have been performed (and for some sources previously has been made by other authors) the lack of sensitive photometry at wavelengths between $\sim$100--500\,$\mu$m still hampers our ability to accurately determine grain properties and size distributions in cold debris disks. This will be addressed with ongoing \emph{Herschel} PACS/SPIRE observations of disks, which are filling the gap between current far-IR and submm measurements, and will detect dust masses down to a few Kuiper-Belt masses for nearby stars \citep{Danchi2010}. With such data, any deviation from the theoretical power law size distribution of observable dust, e.g.\  the ``wavy" size distribution predicted from numerical models of collisional processes in debris disks \citep{Thebault2007,Krivov2006}, can be thoroughly investigated. In addition, ALMA will enable detection and imaging of many faint disks, and permit detailed SED modelling of the very coldest Kuiper-Belt systems.


\subsection{Dust Masses and Characteristic Radial Distances}
For an optically thin disk at submm wavelengths we can make an estimate of the mass of large and cool dust grains (which contain most of the dust mass), providing a lower limit on the total dust mass required to reproduce the measured flux \citep{Hildebrand1983}: 
\begin{equation}
M_{\mathrm{dust}}=\frac{F_{\nu}d^2}{{\kappa_{\nu}}B_{\nu}(T_{\mathrm{dust}})},
\label{eqn:2}
\end{equation}
where $d$ is the distance to the source, and the integrated flux density $F_{\nu}$ at 870\,$\mu$m is assumed to lie on the Rayleigh-Jeans tail of the SED, thus $B_{\nu}$ appropriately expressed as $B_{\nu}=2{\nu}^{2}kT/c^2$. The opacity (or dust mass absorption coefficient), $\kappa_{\nu}$ at submm wavelengths is currently unknown due to uncertainties in the composition and structure of dust grains, but based on results from \citet{Draine1984} the 870\,$\mu$m opacity for micron-sized spherical silicate grains should be $\sim$0.5\,cm$^2$g$^{-1}$, and approximately four times as high for graphite grains at a temperature of 100\,K. Allowing for porous grains would significantly increase the opacity \citep[see][and references therein]{Stognienko1995}, but we adopt $\kappa_{\nu}$=2\,cm$^2$g$^{-1}$ \citep[as in][]{Liseau2003,Nilsson2009}, similar to the commonly chosen 1.7\,cm$^2$g$^{-1}$ \citep[e.g.][]{Zuckerman1993,Dent2000,Najita2005}. The distance $d$ to the source and the derived dust temperature $T_{\mathrm{dust}}$ are two additional parameters with large errors that amplify the inaccuracy of the mass determination. Calculated $M_{\mathrm{dust}}$ is given in Table\,\ref{table:2} with errors based only on the uncertainty in the integrated flux density.

In general, the estimated dust masses fall in the mass range of a few moon masses which is typical for submm detected debris disks at current sensitivities. One exception is HD\,95086 with a dust mass in excess of 76\,$M_{\mathrm{Moon}}$ assuming a second disk component at the upper 45\,K temperature limit.

Another disk parameter that can be obtained from the best-fit $T_{\mathrm{dust}}$ and $\beta$ is the characteristic radial dust distance $R_{\mathrm{dust}}$, which can be found by assuming thermal equilibrium of dust grains \citep[see e.g.][]{Emerson1988}, i.e.\ the emitted (modified blackbody) power
\begin{equation}
P_{\mathrm{e}}=4{\pi}a^{2}F_{\mathrm{dust}}=8h\left(\frac{{\pi}a}{c}\right)^{2}\left(\frac{1}{\nu_{0}}\right)^{\beta}\left(\frac{kT_{\mathrm{dust}}}{h}\right)^{4+\beta}\int\limits_{0}^{\infty}\frac{x^{3+\beta}}{e^{x}-1}dx
\label{eqn:3}
\end{equation}
equals the absorbed power
\begin{equation}
P_{\mathrm{a}}=\frac{R_{*}^{2}{\sigma}T_{\mathrm{e}}^{4}}{R_{\mathrm{dust}}^{2}}{\pi}a^{2}\left(1-A\right).
\label{eqn:4}
\end{equation}
In (\ref{eqn:3}) we made the variable substitution $x=(h{\nu})/(kT_{\mathrm{dust}})$. The integral can be analytically expressed yielding a characteristic radial dust distance given by
\begin{equation}
{R_{\mathrm{dust}}}=\left(\frac{L_{*}\left(1-A\right)c^{2}\nu_{0}^{\beta}}{32\pi^{2}h\left(kT_{\mathrm{dust}}/h\right)^{4+{\beta}}{\zeta}\left(4+{\beta}\right){\Gamma}\left(4+{\beta}\right)}\right)^{1/2},
\label{eqn:5}
\end{equation}
where $L_{*}$ is the luminosity of the star, and $A$ is the albedo of the dust grains. $\zeta\left(s\right)=\Sigma_{n=1}^{\infty}n^{-s}$ is the Riemann zeta function and $\Gamma(n)$ is the Gamma function. Here we also have to assume a $\nu_{0}$ and corresponding grain size $a$ for which the adopted power-law dependence of the opacity is valid. According to the results from \citet{Draine1984} the power-law relation holds for $\lambda_{0}{\lesssim}3$\,$\mu$m, and we choose $a=1.0$\,$\mu$m in our calculation. The albedos of micron-sized dust grains are generally below 0.1, as shown by modelling \citep{Shen2009} and observations of cometary and zodiacal dust \citep{Lasue2006, Reach2003}, thus we use $A=0$, and note that an albedo of almost 0.20 would be required to reduce $R_{\mathrm{dust}}$ with 10\%. The results are presented in the last column of Table\,\ref{table:2}. With the exception of HD\,98800B, for which we derive a characteristic dust distance of $\sim$2\,AU in accordance with e.g.~\citet{Furlan2007,Akeson2007,Prato2001} and consistent with the smallest stable circumbinary orbit \citep{Holman1999}, all detected disks have $R_{\mathrm{dust}}$ comparable to Kuiper-Belt distances or beyond.

The inferred sizes of the detected disks is another reason for our hesitation in regarding the extended flux density distribution around HD\,109573 and HD\,141569 as resolved disks. At distances of 67\,pc and 99\,pc, and sizes of 77\,AU and 110\,AU, respectively, their angular size in comparison with some other disks (e.g.\ the 170\,AU disk around HD\,170773, 36\,AU away) would be smaller (assuming similar grain properties and size distributions), suggesting these other disks should also have been resolved in our observations.

\subsection{Fractional Dust Luminosities}
Since stellar distances, dust opacities and temperature distributions are uncertain, the error in the determined $M_{\mathrm{dust}}$ is actually larger than the ones given in Table\,\ref{table:2} (which was derived from measured flux-density errors), but difficult to quantify. A more independent estimate of the amount of dust in the system can be made by calculating the fractional dust luminosity, $f_{\mathrm{dust}}=L_{\mathrm{dust}}/L_{*}$, from the fitted SEDs. In Table\,\ref{table:2} we have listed fractional dust luminosities found from our SED modelling of detected sources, and values from the literature for undetected sources. Again HD\,98800 stands out, with an exceptionally large $f_{\mathrm{dust}}$, attributed to hot dust belts on the order of $\sim$1\,AU from the binary HD\,98800B \citep{Furlan2007}.

Since all observed stars are nearby the effects of interstellar extinction should be be negligible, which is also confirmed by $B-V$ colour excesses below $\sim$0.1. The only exception is again HD\,152404 (which is however not modelled) with $E_{B-V}=0.28$ . This could be both due to its distance (145 pc) and its T~Tauri nature. The general consequence of uncorrected reddening would be an overestimation of the fractional dust luminosity, and has to be checked for in sources outside the Local Bubble.

%

\subsection{Temporal Evolution of Debris Disks}\label{sec:tempevo}
\begin{figure}[h]
         \resizebox{\hsize}{!}{\includegraphics[width=1.0\textwidth]{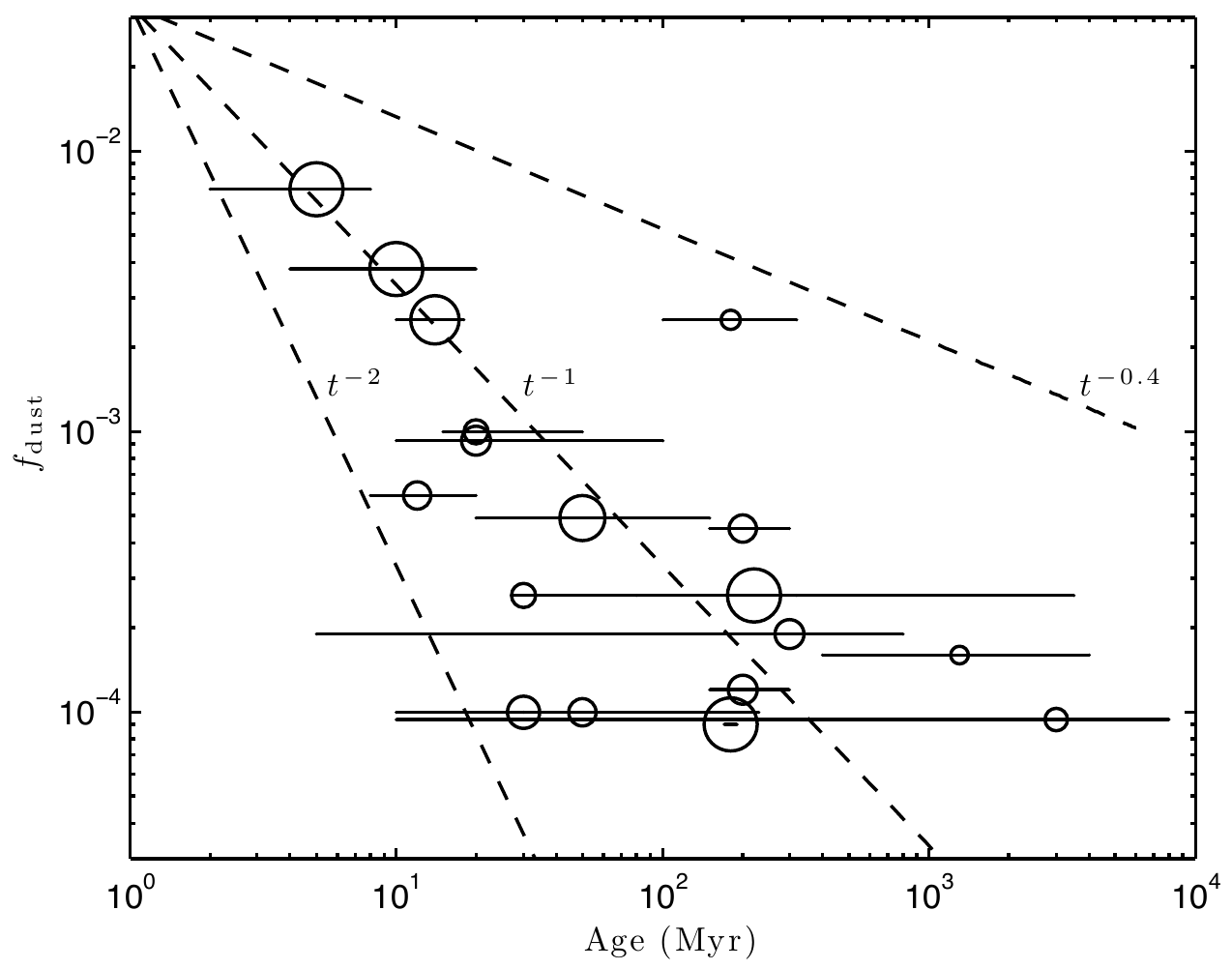}}
	\caption{Fractional dust luminosity plotted against stellar age for the observed sample of stars. For detected sources our derived values have been used, while values from the literature (where available) have been used for undetected ones (see references in Table\,\ref{table:2}). Age references are given in Table\,\ref{table:1}. The size of the circular markers represent the stellar spectral type (diameter increase linearly with earlier spectral type). Dashed lines have been inserted for comparison with the temporal evolution found from previous observations and models.}
         \label{fig:1}
\end{figure}

Both theoretical and previous observational work have indicated that the amount of dust in debris disks, on the average, will decrease from $\sim$10\,Myr as the disk ages. This is to be expected, due to the combined effect of collisional grinding producing small grains that are blown out by radiation pressure, Poynting-Robertson drag, sublimation and photosputtering of ices \citep[e.g.][]{Wyatt2007,Wyatt2008}, and is the general behavior confirmed in debris disk surveys \citep[e.g][]{Najita2005,Su2006}. \citet{Spangler2001} found  $f_{\mathrm{dust}}$\,${\propto}$\,$t^{-\alpha}$ with $\alpha=2$ from ISO observations of clusters, but these results were not confirmed by later studies that found significantly lower $\alpha$, e.g.\ \citet{Liu2004} derived ${\alpha}$\,=\,0.5--1.0 and \citet{Su2006} ${\alpha}=0.6$, from \emph{Spitzer} and JCMT observations, respectively. The main uncertainty lies in the inaccuracy of current age estimates of individual stars, as discussed in Section\,\ref{sec:sample}, and possible variations in disk evolution depending on stellar spectral types. \citet{Nilsson2009} compared the average fractional dust luminosity observed in the $\sim$12\,Myr {$\beta$} Pictoris Moving Group with the $\sim$100\,Myr Pleiades cluster \citep{Greaves2009} and obtained $\alpha>0.8$. Interestingly, these values are higher than those predicted from models of collisional dust evolution, which predict a slower decline, with $\alpha$ as low as 0.4 \citep{Wyatt2007,Lohne2008}. 
In Fig.\,\ref{fig:1} we have plotted $f_{\mathrm{dust}}$ versus age for the observed sample of stars in this study, with the diameter of the circular marker scaled linearly with stellar spectral type (increasing for earlier spectral types). Due to the large age uncertainties (represented by horizontal error bars) we refrain from making a fit, and instead only display dashed lines showing the expected decline for three different values of $\alpha$. We conclude that the temporal evolution of the fractional dust luminosity determined from our observations is in agreement with \citet{Nilsson2009}, and $f_{\mathrm{dust}}$ could perhaps decline as fast as the $t^{-2}$ found by \citet{Spangler2001}, although $\alpha<0.8$ can not be excluded due to large error bars and small stellar sample. This trend will be better constrained after the completion of the ongoing APEX/LABOCA Large Programme.

\begin{figure}[h]
         \resizebox{\hsize}{!}{\includegraphics[width=1.0\textwidth]{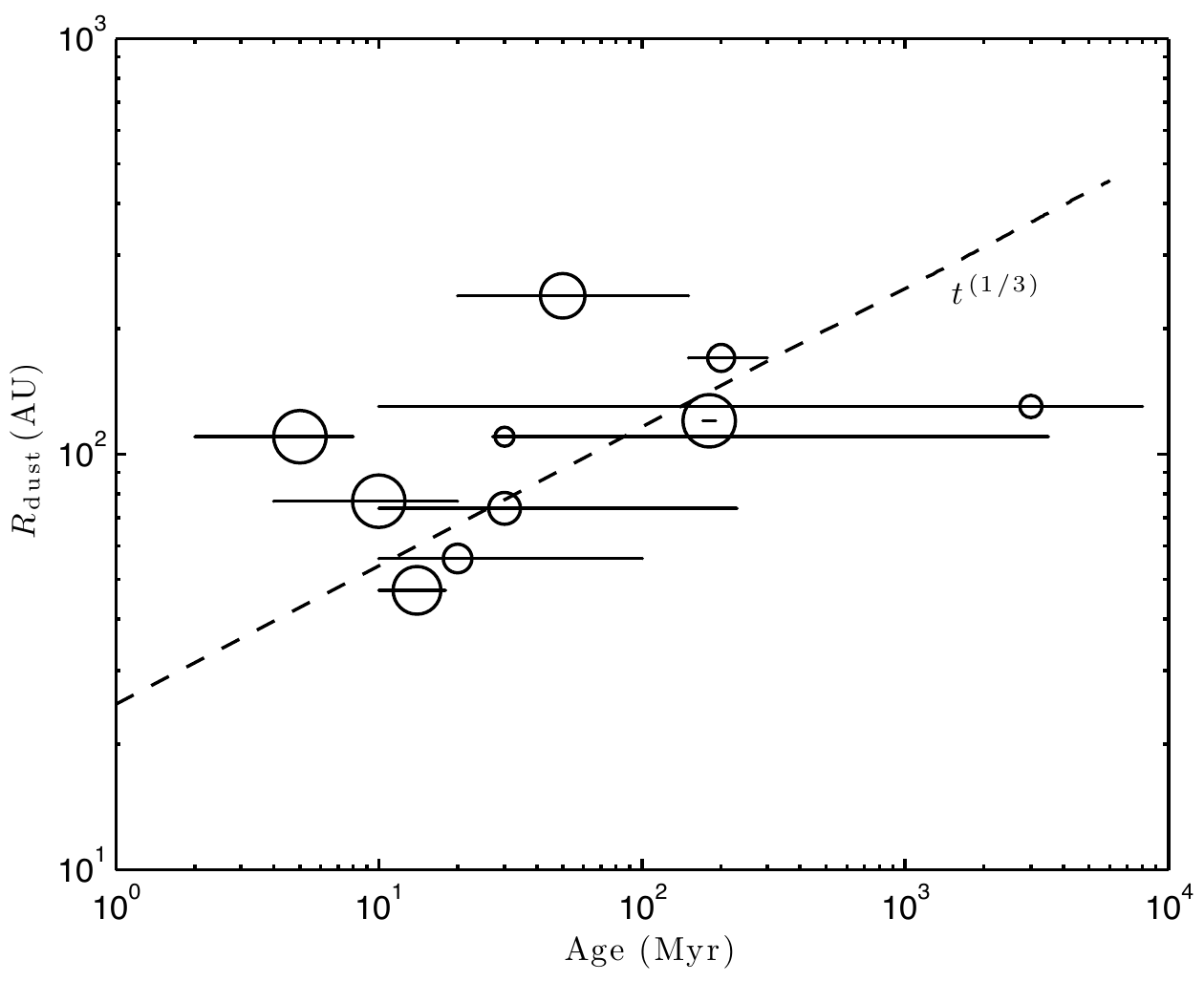}}
	\caption{Characteristic radial dust distance for detected submm disks plotted against stellar age. Age references can be found in Table\,\ref{table:1}. The dashed line shows the expected slope for the temporal evolution of dust in a planetesimal disks dominated by self-stirring \citep{Kenyon2008,Kenyon2004a}.}
         \label{fig:2}
\end{figure}

Note that this issue is an important one because the value of $\alpha$ should be an indication of the dominant process for dust removal: $\alpha$\,$\geq$\,1 values are in principle only possible for systems at steady-state where Poynting-Robertson drag dominates, while $\alpha$\,$\leq$\,1 corresponds to collision dominated systems at steady state \citep{Dominik2003}. However, the problem could be more complex if the collisional evolution of the system is controlled by self-stirring, triggered by the formation of large, 1000--2000\,km sized embryos. In this case, the system can brighten from an early low state before later reaching the steady state collisional decline \citep{Kenyon2008,Kenyon2004a, Wyatt2008}. An interesting characteristic of self-stirred disks is that the peak of dust production, which coincides with the formation of planetary embryos, should be reached at different times depending on the radial distance to the star. This dust production peak should thus propagate outward with time, leading to a progressive increase of the observed characteristic dust distance $R_{\mathrm{dust}}$ with age as $t^{1/3}$. We have tried to identify this evolutionary trend by plotting $R_{\mathrm{dust}}$ of our detected submm disks versus age (in Fig.\,\ref{fig:2}). In contrast to \citet{Najita2005}, who found no apparent correlation for a sample of 14 submm detected disks, we do identify an increase in characteristic radial dust distance with increasing age, although it is again difficult to make any firm conclusion regarding the power-law exponent. These tentative results will also have to be confirmed with additional detections in ongoing observations.

For both the fractional dust luminosity and the characteristic radial dust distance there seems to be no obvious dependence on stellar spectral type.

\addtocounter{table}{1}

\section{Conclusions}
The most important results from our 870\,$\mu$m observations of 22 exo-Kuiper-Belt candidates can be summarised as follows:
   \begin{itemize}
      \item[\textbullet] Out of the observed sample of far-IR excess stars we detected $\sim$45\% (10 out of 22) with at least a 3$\sigma$ significance. Five of the detected debris disks (HD\,95086, HD\,131835, HD\,161868, HD\,170773, and HD\,207129) have previously never been seen at submm wavelengths. These findings increase the currently known exo-Kuiper-Belts (inferred from cold extended dust disks detected in the submm) from 29 to 34. The sensitivity of current submm observations with LABOCA at APEX and SCUBA at JCMT are reaching the background extragalactic confusion limit, as evident by the abundance of significant off-center flux density peaks in acquired submm maps. Future ALMA observations will be essential to discriminate between resolved disks and background submm galaxies. 
      \item[\textbullet] We perform SED modelling of available photometric data using a simple modified blackbody to make a $\chi^{2}$ minimization fit to the disk excess emission in order to determine the dust temperature $T_{\mathrm{dust}}$ and power-law exponent $\beta$ of the opacity law. Resulting $\beta$ parameters are all between 0.1 and 0.8, suggesting grains significantly larger than those in the ISM, in agreement with previous results for debris disks \citep[e.g.][]{Najita2005,Nilsson2009,Roccatagliata2009}. However, it should be noted that the presence of colder secondary dust components in addition to the fitted disk could be responsible for some abnormally low $\beta$ values. The measured 870\,$\mu$m flux of HD\,95086 undeniably requires a second disk component. 
      \item[\textbullet] Besides finding the dust mass of detected sources (together with upper limits on undetected ones) derived from the integrated 870\,$\mu$m flux density, we use the best-fit SED parameters to calculate the fractional dust luminosity $f_{\mathrm{dust}}$ and the characteristic radial dust distance $R_{\mathrm{dust}}$ (for 1$\mu$m sized grains) of the disks. We present plots of the temporal evolution of $f_{\mathrm{dust}}$ and $R_{\mathrm{dust}}$ based on these values together with previously estimated stellar ages, and find that our results are consistent with a $f_{\mathrm{dust}}$\,$\propto$\,$t^{-\alpha}$, with $\alpha$\,$\sim$\,0.8--2.0, as well as tentatively with the $R_{\mathrm{dust}}$\,$\propto$\,$t^{1/3}$ relation predicted in debris disks collisional evolution models by \citet{Kenyon2004a}. 
   \end{itemize}

\begin{acknowledgements}
This research was supported by financial contributions from Stockholm Astrobiology Graduate School, and the International Space Science Institute (ISSI) in Bern, Switzerland (``Exozodiacal Dust Disks and Darwin'' working group, \texttt{http://www.issibern.ch/teams/exodust/}). A.B. was funded by the \textit{Swedish National Space Board} (contract 84/08:1).
\end{acknowledgements}

\bibliographystyle{aa}
\bibliography{myref_largeprog}


\begin{appendix}

\section{Spatially Resolved Disks or Chance Alignment of Background Galaxies?}\label{sec:appena}
Two sources appear spatially resolved in our 870\,$\mu$m maps, both of them are multiple systems which have been previously imaged at shorter wavelengths. Here we compare the resolved structures at submm wavelengths with that observed in scattered optical light and IR emission, and discuss the indications for and against them actually being resolved disks. The position angle (PA) was found by fitting a 2-D Gaussian profile to the extended submm emission.
\subsection{HD\,109573 (HR\,4796)}
The highly inclined debris disk around HD\,109573A (also known as HR\,4796A) was first imaged in the mid-IR by \citet{Koerner1998} and \citet{Jayawardhana1998}, followed by optical and near-IR coronagraphic imaging by \citet{Schneider1999,Schneider2009}. These observations revealed a 17\,AU wide dust ring with $\sim$70\,AU radius surrounding the primary A0V star. In our map (Fig.\,\ref{fig:A1f}), the peak flux density of 19.1$\pm$3.1\,mJy/beam is centered on the position of the primary, with the secondary M2.5 star located 7$\farcs$7 ($\sim$2 pixels) to the southwest. Surprisingly, the PA is 129$\degr$, compared to 27$\degr$ for the ring imaged by \citet{Schneider2009}, i.e.\ a nearly orthogonally oriented extension. Although the accuracy of our determined PA is hard to estimate due to the influence of possible lower-sigma noise features, the flux distribution is undoubtedly extended in the southeast direction, maybe with a warp toward the East. The 3$\sigma$ contour reaches $\gtrsim$2000\,AU projected radial distance on that side of the central star. However, a chance alignment with a bright background galaxy to the southeast is probable, for several reasons. Firstly, an outer disk orthogonally oriented to the inner one seems highly implausible or even unphysical. In addition, a data reduction employing only half-beam smoothing seem to imply a separate 3$\sigma$ peak in the southeast flux extension. A second galaxy to explain the small northwest extension is conceivable, but it could also be a noise feature. The formal likelihood of a source falling within some 25$\arcsec$ distance from the star and making it appear extended is about 4\%.
\subsection{HD\,141569}
This triple system has been extensively studied since the \emph{Hubble Space Telescope} (HST) discovery of a large disk surrounding the primary A0Ve star HD\,141569A \citep{Weinberger1999,Augereau1999b}. The disk has been imaged in thermal mid-IR emission \citep{Fisher2000}, but up until now never resolved at longer wavelengths. We find a clear asymmetric elongation of the disk with PA of 10$\degr$ to the southwest of the source-centered 10.3$\pm$2.3\,mJy/beam peak flux (Fig.\,\ref{fig:A1h}). This can be compared to the two most detailed studies of the disk structure so far, employing coronagraphic HST ACS and STIS observations, respectively \citep{Clampin2003,Mouillet2001}, which revealed an inner ($\lesssim$175\,AU) clearing surrounded by two ring-like structures - a thin ($\sim$50\,AU) belt at $\sim$200\,AU and a wider ($\sim$100\,AU) belt at $\sim$350\,AU - oriented with a semi-major axis roughly in the North-South direction and connected by faint spiral structures. \citet{Clampin2003} also found additional spiral arcs extending from the northeast (out to $\sim$1200\,AU) and the southwest (toward the binary HD\,141569BC, located $\sim$8$\arcsec$ northwest of HD\,141569A. The reason for the highly asymmetric brightness distribution in the optical, with respect both to semi-major and semi-minor axis, could be a non-axisymmetric distribution of grains and/or anisotropic scattering by grains \citep{Mouillet2001}. Suspected gravitational perturbation by a massive body in the disk, and effects from the binary companion has been extensively studied by e.g.\ \citet{Augereau2004,Quillen2005,Wyatt2005}, while \citet{Ardila2005} and \citet{Reche2009} thoroughly investigated the case of a gravitationally unbound binary passing by (with and without planets in the disk), but did not manage to explain all observed features. It is possible that we are seeing a colder dust population on very eccentric orbits, perhaps shaped by the dynamical interaction with the binary (either in a flyby or a triple system scenario), however, dynamical modelling of such a scenario is beyond the scope of the present paper. The extreme brightness asymmetry with respect to the semi-minor axis, on the other hand, does not make us confident that the extended emission originates from the dust disk, and could instead be attributed to a background galaxy located southwest of HD\,141569. Also in this case a re-reduction applying only half-beam smoothing seem to hint at a separate 3$\sigma$ peak, only 8$\arcsec$ to the southeast of the star (which would be a 0.5\% chance alignment).

\section{List of submm detected exo-Kuiper-Belts}\label{sec:appenb}
\begin{table*}
 \caption{\label{table:3} Submm and mm photometry of all debris disks that have been detected at submm and longer wavelengths.}
\centering
\begin{tabular}{l c c c c}
\hline\hline
Star & Other Name &  $\lambda$ & $F_{\nu}$ & Reference\\
 &  & ($\mu$m) & (Jy) &  \\
 \hline
HD\,105 &  & 870 & $10.7\pm5.9$\tablefootmark{a} & 1\\

HD\,377 &  & 1200 & $4.0\pm1.0$ & 11\\

HD\,8907 &  & 1200 & $3.2\pm0.9$ & 11\\

HD\,14055 & $\gamma$~Tri & 850 & $5.5\pm1.8$ & 3\\

HD\,15115 &  & 850 & $4.9\pm1.6$ & 3\\
 
HD\,17206 & ${\tau}^{1}$~Eri & 1300 & $20.7\pm3.9$ & 6\\

HD\,21997 &  & 870 & $8.3\pm2.3$ & 3\\ 

 &  & 870 & $17.6\pm8.0$ & 1\\

HD\,22049 & ${\epsilon}$~Eri & 450 & $225\pm10$ & 12\\

 &  & 850 & $40.0\pm1.5$ & 12\\

 &  & 1300 & $24.2\pm3.4$ & 6\\
 
HD\,30447 &  & 870 & $6.9\pm5.0$\tablefootmark{a} & 1\\

HD\,32297 &  & 1300 & $5.1\pm1.1$ & 10\\

HD\,38393 &  & 850 & $2.4\pm1.0$ & 12\\

HD\,39060 & ${\beta}$~Pic & 850 & $58.3\pm6.5$ & 4\\

 &  & 870 & $63.6\pm6.7$ & 2\\
 
 &  & 1200 & $24.3\pm3.0$ & 5\\

 &  & 1300 & $24.9\pm2.6$ & 6\\

HD\,39944 &  & 1350 & $3.7\pm0.9$ & 14\\
 
HD\,48682 & 56~Aur & 850 & $5.5\pm1.1$ & 12\\

HD\,61005 &  & 350 & $95\pm12$ & 11\\
 
HD\,95086 &  & 870 & $41.3\pm18.4$ & 1\\

HD\,98800\tablefootmark{b} & TV~Crt & 870 & $33.6\pm8.4$ & 1\\

 &  & 1350 & $36.8\pm4.2$ & 14\\

 &  & 2000 & $24.8\pm3.4$ & 14\\

HD\,104860 &  & 350 & $50.1\pm9.3$ & 11\\
 
 &  & 1200 & $4.4\pm1.1$ & 11\\
 
HD\,107146 &  & 350 & $319\pm6$ & 11\\
 
 &  & 450 & $130\pm40$ & 7\\
 
 &  & 850 & $20\pm4$ & 7\\
 
HD\,109085 & ${\eta}$~Cor & 850 & $7.5\pm1.2$ & 12\\
 
HD\,109573 & HR\,4796 & 450 & $180\pm150$ & 12\\
 
 &  & 850 & $19.1\pm3.4$ & 8, 12\\
 
 &  & 870 & $21.5\pm6.6$ & 1\\
 
HD\,123160 &  & 850 & $13\pm4.3$ & 14\\
 
 &  & 1350 & $4.7\pm0.9$ & 14\\ 

HD\,128167 & $\sigma$~Boo & 850 & $6.2\pm1.7$ & 12\\

HD\,131835 &  & 870 & $8.5\pm4.4$ & 1\\
 
HD\,141569\tablefootmark{b} &  & 450 & $66\pm23$ & 12\\

 &  & 850 & $14\pm1$ & 12\\
 
 &  & 870 & $12.6\pm4.6$ & 1\\

 &  & 1300 & $9.39\pm3.72$ & 13\\
 
 &  & 1350 & $5.3\pm1.1$ & 14\\

HD\,152404\tablefootmark{b} & AK~Sco & 450 & $242\pm32$ & 15\\

 &  & 800 & $83\pm9$ & 15\\
 
 &  & 870 & $42.9\pm9.8$ & 1\\

 &  & 1100 & $36\pm4$ & 15\\

HD\,161868 & $\gamma$~Oph & 870 & $12.8\pm5.2$ & 1\\

HD\,170773 &  & 870 & $18.0\pm5.4$ & 1\\

HD\,172167 & Vega & 850 & $45.7\pm5.4$ & 4, 12\\

 &  & 1300 & $11.4\pm1.7$ & 16\\

HD\,181327 &  & 870 & $51.7\pm6.2$ & 2\\
 
HD\,191089 &  & 350 & $54\pm15$ & 11\\

HD\,195627 &  & 870 & $13.0\pm7.1$\tablefootmark{a} & 1\\
 
HD\,197481 & AU~Mic & 850 & $14.4\pm1.8$ & 9\\

HD\,207129 &  & 870 & $5.1\pm2.7$ & 1\\

HD\,216956 & Fomalhaut & 450 & $595\pm35$ & 12\\

 &  & 850 & $97\pm5$ & 12\\

 &  & 850 & $81\pm7.2$ & 4\\
 
HD\,218396 & V342~Peg & 850 & $10.3\pm1.8$ & 3\\

HIP\,23200 & V1005~Ori & 850 & $4.8\pm1.2$ & 9\\
\hline
\end{tabular}
\tablebib{
(1)~This paper; (2) \citet{Nilsson2009}; (3) \citet{Williams2006}; (4) \citet{Holland1998}; 
(5) \citet{Liseau2003}; (6) \citet{Chini1991}; (7) \citet{Williams2004}; (8) \citet{Greaves2000b};
(9) \citet{Liu2004}; (10) \citet{Maness2008}; (11) \citet{Roccatagliata2009}; (12) \citet{Sheret2004};
(13) \citet{Walker1995}; (14) \citet{Sylvester2001}; (15) \citet{Jensen1996}; (16) \citet{Wilner2002}.
}
\tablefoot{
\tablefoottext{a}{Marginally detected.}
\tablefoottext{b}{Debris disk status disputed.}
}
\end{table*}

\end{appendix}

\end{document}